\documentclass[%
 aip,
 sd,%
 amsmath,amssymb,
 preprint,%
]{revtex4-1}

\usepackage{graphicx}
\usepackage{dcolumn}
\usepackage{bm}

\begin{document}

\preprint{AIP/123-QED}

\title{Polarity tuning of spin-orbit-induced spin splitting in two-dimensional transition metal dichalcogenides}

\author{Moh. Adhib Ulil Absor}
\email{adib@ugm.ac.id} 
\affiliation{Department of Physics, Universitas Gadjah Mada BLS 21 Yogyakarta Indonesia.}%
\author{Iman Santosa}
\affiliation{Department of Physics, Universitas Gadjah Mada BLS 21 Yogyakarta Indonesia.}%

\author{Harsojo}
\affiliation{Department of Physics, Universitas Gadjah Mada BLS 21 Yogyakarta Indonesia.}%

\author{Kamsul Abraha}
\affiliation{Department of Physics, Universitas Gadjah Mada BLS 21 Yogyakarta Indonesia.}%

\author{Hiroki Kotaka}
\affiliation{Elements Strategy Initiative for Catalysts and Batteries (ESICB), Kyoto University, Kyoto 615-8520, Japan
}%

\author{Fumiyuki Ishii}%
\affiliation{Faculty of Mathematics and Physics Institute of Science and Engineering Kanazawa University 920-1192 Kanazawa Japan.}%

\author{Mineo Saito}
\affiliation{Faculty of Mathematics and Physics Institute of Science and Engineering Kanazawa University 920-1192 Kanazawa Japan.}%

\date{\today}

\begin{abstract}
The established spin splitting in monolayer (ML) of transition metal dichalcogenides (TMDs) that is caused by inversion symmetry breaking is dictated by mirror symmetry operations to exhibit fully out-of-plane direction of spin polarization. Through first-principles density functional theory calculations, we show that polarity-induced mirror symmetry breaking leads to sizable spin splitting having in-plane spin polarization. These splittings are effectively controlled by tuning the polarity using biaxial strain. Furthermore, the admixtures of the out-of-plane and in-plane spin-polarized states in the strained polar systems are identified, which is expected to influence the spin relaxation through the Dyakonov-Perel mechanism. Our study clarified that the polarity-induced mirror symmetry breaking plays an important role in controlling the spin splitting and spin relaxation in the TMDs ML, which is useful for designing future spintronic devices. 
\end{abstract}

\pacs{Valid PACS appear here}
\maketitle

\section{INTRODUCTION}

Exploration of spin-orbit coupled systems is now at the heart of the growing research field of spintronics that focuses on the manipulation of non-equilibrium material properties using spin-orbit coupling (SOC). The SOC is a relativistic interaction arising from electrons movement in the nuclear electric field, which allows for generation and manipulation of electron spin \cite {Kato}. Current-induced spin polarization \cite {Kuhlen} and the spin Hall effect \cite {Qi} are important examples of spintronics phenomena where the SOC plays an important role. For spintronics device operation, semiconducting structures are promising because of their manipulability under gate voltages \cite {Datta,NittaA}. However, a stable two-dimensional (2D) thin film is highly desirable, which is advantageous for circuit integration. Here, some attentions are given for the 2D materials such as graphene \cite {Novoselov} and their analogs like silicene and germanene \cite {Cahangirov} due to their exotic properties such as high carrier mobility and long spin lifetime \cite {Min,LiuC,Kane}. However, the weak SOC effect in these materials \cite {Min,LiuC} may limit their functionality for spintronic applications.

Recently, monolayer (ML) of transition metal dichalcogenides (TMDs), a new class of the 2D materials, has attracted much attention because of their extraordinary properties, especially the exotic spin-valley coupled electronic structures that promise future spintronic and valleytronic applications \cite {Xiao,Xu,Yuan,Zhu,Latzke,Liu_Bin,Absor4,Chu}. The bulk phase of the TMDs systems is characterized by an inversion symmetric of 2$H$ $MX_{2}$ stacking orders with space group $D_{6h}$\cite{Bromley}. In the ML phase, the inversion symmetry is broken, leading to the fact that the symmetry reduces to be $D_{3h}$ \cite{Zhu,Absor4}. This inversion symmetry breaking together with strong SOC in the 5$d$ orbitals of transition metals atoms give rise to large spin splitting \cite {Kosminder,Zhu,Yuan,Liu_Bin,Absor4}, which plays an important role in inducing some of interesting phenomena such as spin Hall effect \cite{Cazalilla,Ma}, spin- and valley-dependent selection rule for optical transitions \cite{Chu}, and magnetoelectric effect \cite {Gong}. Furthermore, in the $D_{3h}$ point group symmetry, mirror symmetry operation on the surface plane of the $MX_{2}$ ML suppresses the spin splitting to exhibit fully out-of-plane spin polarization, which is believed to be responsible for inducing strongly enhanced spin relaxation through Dyakonov-Perel mechanism\cite {Zhu,Absor4,Schmidt,L_Yang}. Previous experimental studies have confirmed that long-lived spin relaxation and spin coherence of electrons have been reported on various $MX_{2}$ ML such as MoS$_{2}$ ML \cite {Schmidt,L_Yang} and WS$_{2}$ ML \cite {L_Yang}.

Because the mirror symmetry in the surface plane of the $MX_{2}$ ML plays a significant role in controlling the spin splitting and spin-polarized states, new electronic properties are expected to appear by breaking this mirror symmetry. Such situation is achieved by introducing the polar structures $MXY$ \cite {Cheng,Defo}, in which the polarity that is induced by out-of-plane distance difference between transition metal ($M$) and chalcogen ($X$,$Y$) atoms breaks the mirror symmetry in the surface plane. Experimentally, it is possible to create such polar structures by growth on the polar substrate by recently developed molecular beam epitaxial (MBE) technique \cite {Xenogiannopoulou, Aretouli}. This is supported by the fact that the stability of various polar $MXY$ ML structures such as the polar WSSe and MoSSe MLs has recently been reported \cite {Cheng,Defo}. However, due to the structural-dependent of the polarity, controlled changes in the polar structure, for example by applying strain, effectively tunes the polarity, which is expected to induce useful properties for spintronics.

In this paper, we perform first-principles density functional theory calculations to clarify the polarity-strain dependent on the electronic properties of the TMDs ML. We find that in addition to the established spin splitting having fully-out-of-plane spin polarization, a sizable spin splitting exhibiting in-plane spin polarization is observed in the polar $MXY$ ML. These splittings are found to be effectively controlled by tuning the polarity, which is achieved by applying biaxial strain. The origin of the spin splitting and spin-polarized states is analyzed on the basis of symmetry arguments combined with orbital hybridization analyses. Furthermore, the admixtures of the out-of-plane and in-plane spin-polarized states in the strained polar systems are identified, and their implications to the spin relaxation are discussed. Finally, the possible applications of our systems for spintronics are discussed.

\section{Model and Computational Details}

Similar to the case of the non-polar $MX_{2}$, crystal structure of the polar $MXY$ consists of $X$-$M$-$Y$ slabs weakly bonded by van der Waals interaction \cite {Bromley}. Here, an intermediate layer of hexagonally arranged the transition metal atoms ($M$) is sandwiched between two layers of the chalcogenide atoms ($X$,$Y$) through strong ionic-covalent bonds forming a trigonal prismatic arrangement. In the bulk phase, both the non-polar $MX_{2}$ and polar $MXY$ structures have an inversion symmetric of a 2$H$ stacking order with a space group of $D_{6h}$. However, this inversion symmetry is broken in the monolayer (ML) phase. In the case of the non-polar $MX_{2}$ ML, its symmetry reduces to be $D_{3h}$. This symmetry consists of a threefold rotation $C_{3}$ around the trigonal $z$ axis and two mirror symmetry operations with respect to the $x-y$ plane ($M_{x-y}$) and the $y-z$ plane ($M_{y-z}$) [Fig. 1(a)]. On the contrary, the mirror symmetry $M_{x-y}$ is broken in the case of the polar $MXY$ ML, leading to the fact that the symmetry becomes $C_{3v}$ [Fig. 1(b)]. The broken of mirror symmetry $M_{x-y}$ in the polar $MXY$ ML is induced by the polarity originated from the out-of-plane interlayer distance difference, $\Delta d_{\bot}$. Here, $\Delta d_{\bot}$ is defined as $\Delta d_{\bot}=\left|d_{\bot (M-Y)}-d_{\bot (M-X)}\right|$, where $d_{\bot (M-X)}$ and $d_{\bot (M-Y)}$ are the distance between $M$ atoms and $X$ or $Y$ atoms in the out-of-plane direction. Because the physics in the polar $MXY$ and non-polar $MX_{2}$ ML systems are essentially the same for the group- VI transition metal dichalcogenides (TMDs), we here choose WSSe and WS$_{2}$ MLs as an example of the polar $MXY$ and non-polar $MX_{2}$ MLs, respectively.   

We performed first-principles electronic structure calculations based on the density functional theory (DFT) within the generalized gradient approximation (GGA) \cite {Perdew} using the OpenMX code \cite{Openmx}. We used norm-conserving pseudopotentials \cite {Troullier}, and the wave functions are expanded by the linear combination of multiple pseudoatomic orbitals (LCPAOs) generated using a confinement scheme \cite{Ozaki,Ozakikino}. The orbitals are specified by W7.0-$s^{2}p^{2}d^{2}f^{1}$, S9.0-$s^{2}p^{2}d^{1}$, and Se9.0-$s^{2}p^{2}d^{1}$, which means that the cutoff radii are 7.0, 9.0, and 9.0 bohr for the W, S, and Se atoms, respectively, in the confinement scheme \cite{Ozaki,Ozakikino}. For the W atoms, two primitive orbitals expand the $s$, $p$, and $d$ orbitals, and one primitive orbital expands the $f$ orbital. On the other hand, for the S and Se atoms, two primitive orbitals expand the $s$ and $p$ orbitals, and one primitive orbital expands $d$ orbital. The SOC was included in our DFT calculations, and the spin textures in $k$-space were calculated using the $k$-space spin density matrix of the spinor wave function \cite{Kotaka, Absor1,Absor2,Absor3,Absor4}.

\begin{figure}
	\centering
		\includegraphics[width=0.9\textwidth]{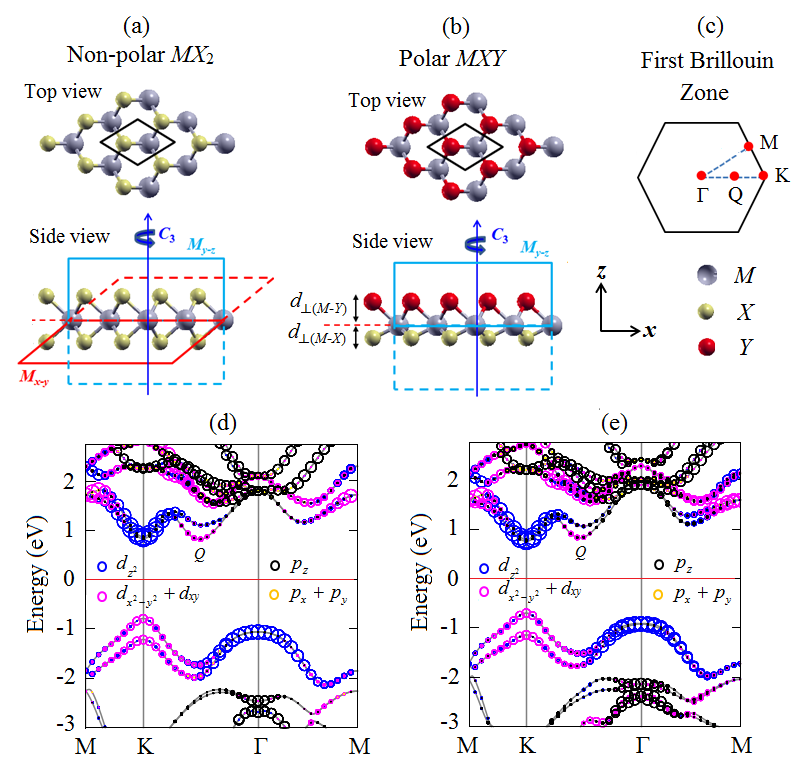}
	\caption{Top and side views of (a) the non-polar $MX_{2}$ and (b) the polar $MXY$ monolayers (MLs) structures.  These structures are characterized by a threefold rotation $C_{3}$ around the trigonal $z$ axis and two mirror symmetry operations with respect to the $x-y$ plane ($M_{x-y}$) and to the $y-z$ plane ($M_{y-z}$). Out-of-plane distance between transition metal ($M$) and chalcogen atoms ($X$,$Y$) [$d_{\bot (M-Y)}$, $d_{\bot (M-X)}$] is indicated. First Brillouin zone, which is specified by the high symmetry points ( $\Gamma$, $K$, and $M$ points ) is shown. Here, the points located along the $\Gamma$-$K$ line, namely the $Q$ point is indicated. Orbital-resolved of the electronic band structures for (c) the non-polar WS$_2$ ML and (d) the polar WSSe ML. The radius of circles reflects the magnitudes of spectral weight of the particular orbitals to the band. The calculations are performed with inclusion the effect of the spin-orbit coupling.}
	\label{figure:Figure1}
\end{figure}

Two dimensional structures of the non-polar $MX_{2}$ and polar $MXY$ MLs are modeled as a periodic slab with a sufficiently large vacuum layer (25 \AA). The use of the large vacum layer in the present system is to ensure that the electron density of the material tails off to zero in the vacuum and to avoid interaction between adjacent layers along the direction perpendicular to the surface plane. The geometries were fully relaxed until the force acting on each atom was less than 1 meV/\AA. Here, we find that the optimized in-plane lattice constant is 3.24 \AA\ for the case of the polar WSSe ML, which is larger than that of the non-polar WS$_2$ ML (3.18 \AA), but it is in a good agreement with previous results [3.24 \AA\ \cite{Cheng} to 3.25 \AA\ \cite{Defo}]. Furthermore, we characterize the degree of the polarity in our system by evaluating $\Delta d_{\bot}$. In the case of the non-polar WS$_{2}$ ML, we find that the calculated value of $\Delta d_{\bot}$ is zero, indicating that this structure is symmetric, thus, justifying the non-polarity of this structures. On the contrary, in the case of the polar WSSe ML, it is found that the calculated value of $\Delta d_{\bot}$ is 0.154 \AA, showing that this system is polar. Because of the structural-dependent of $\Delta d_{\bot}$, it is expected that the polarity can be controlled by applying strain.
   
\section{RESULT AND DISCUSSION}

\subsection{Electronic structures and characteristic of the spin splitting}

To investigate the effect of the polarity on the electronic properties of the TMDs ML, we show in Figs. 1(d) and 1(e) orbital-resolved of electronic band structures calculated on the first Brillouin zone [Fig. 1(c)]. In the case of the non-polar WS$_{2}$ ML, we observe two local maxima in the valence band maximum (VBM) located on the $K$ and $\Gamma$ points, which are predominately filled by $d_{{x^2}-{y^2}}+d_{xy}$ and $d_{z^2}$ orbitals, respectively [Fig. 1(d)]. On the other hand, in the conduction band minimum (CBM), we identify two local minima with close in energy located on the $K$ point and midway between the $\Gamma$ and $K$ points, namely the $Q$ points. These local minima at $Q$ and $K$ points are mainly originated from the $d_{{x^2}-{y^2}}+d_{xy}$ and $d_{z^2}$ orbitals, respectively [Fig. 1(d)]. Since the VBM and CBM are centered at the $K$ point, a direct band gap with an energy gap of 1.68 eV is observed, which is in good agreement with previous calculations \cite{Absor4,Guzman}. Similar to the case of the non-polar WS$_{2}$ ML, the direct band gap is also identified in the case of the polar WSSe ML [Fig. 1(e)]. However, the energy gap of the polar WSSe ML (1.55 eV) is smaller than that of the non-polar WS$_{2}$ ML. Due to the out-of-plane crystal asymmetry in the polar WSSe ML [Fig. 1(b)], hybridization between the out-of-plane orbitals [$d_{z^2}$, $p_{z}$] enhances, resulting in that energy level of the $K$ point in the CBM shifts to be lower than that observed on the non-polar WS$_{2}$ ML, thus induces lowering the band gap. These features of electronic band structures are consistent with recent observation of the electronic properties across the TMDs family \cite {Zhu,Latzke,Liu_Bin,Absor4,Cheng}. 

\begin{table}[h!]
\caption{The calculated value of spin splitting in different high symmetry points in the first Brillouin zone for the VBM and CBM.  Here, $\Delta E_{K,\texttt{VBM}}$ denotes the spin splitting at the $K$ point in the VBM, while $\Delta E_{K,\texttt{CBM}}$ and $\Delta E_{Q,\texttt{CBM}}$ denote the spin splitting at the $K$ and $Q$ points in the CBM, respectively. A comparison with previous results are also shown.} 
\centering 
\begin{tabular}{c c c c c} 
\hline\hline 
Monolayer (ML) & $\Delta E_{K,\texttt{VBM}}$ (eV) & $\Delta E_{K,\texttt{CBM}}$ (eV)  & $\Delta E_{Q,\texttt{CBM}}$ (eV) & Reference \\ 
\hline 
Non-polar WS$_{2}$ ML & 0.43 & 0.03 & 0.33 & This work \\ 
         & 0.43 &   -   &   -   &  Ref. \cite{Zhu} \\
         & 0.43 & 0.03 & 0.33 &  Ref. \cite{Absor4} \\
				 & 0.41-0.47 & - & - &  Ref. \cite{Latzke} \\
				 & 0.43 & 0.03 & - &  Ref. \cite{Liu_Bin} \\
				 & 0.43 & 0.03 & 0.26 &  Ref. \cite{AndorB} \\
  Polar WSSe ML  & 0.25 & 0.04 & 0.20 & This work \\ 
         & 0.44 & 0.03 &-& Ref. \cite{Cheng}\\ 
\hline\hline 
\end{tabular}
\label{table:Table 1} 
\end{table}

Turning the SOC, a spin splitting in the electronic band structures is established in both the non-polar and polar TMDs ML due to the absence of inversion symmetry [Fig. 2]. In the case of the non-polar WS$_{2}$ ML, the exsistence of the mirror symmetry in the surface plane ($M_{x-y}$) suppreses the spin splitting in the band structures except for $\vec{k}$ point along the $\Gamma$-$M$ direction. On the contrary, the mirror symmetry $M_{x-y}$ is broken in the case of the polar WSSe ML, leading to the fact that the spin degeneracy of the bands along the $\Gamma$-$M$ direction is lifted. It is noted here that due to time reversability, the spin degeneracy is visible in the $\Gamma$ and $M$ points. However, at the $K$ and $Q$ points, the time reversal symmetry is broken, inducing Zeeman-like spin splitting \cite {Yuan}. 

\begin{figure}
	\centering
		\includegraphics[width=0.8\textwidth]{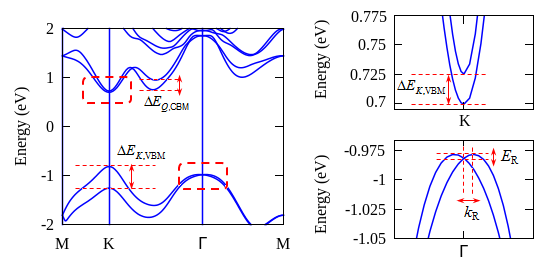}
	\caption{The spin split bands in the polar WSSe ML are given. The spin splitting in the $K$ and $Q$ are indicated by the red lines-arrows. The spin splitting around the $\Gamma$ point known as Rashba splitting is higlighted, which is characterized by the Rashba energy ($E_{R}$), and momentum offset ($k_{R}$).}
	\label{figure:Figure2}
\end{figure}

The calculated values of the spin splitting in different high symmetry points in the first Brillouin zone for the VBM and CBM are summarized in Table I. Consistent with previous studies \cite {Zhu,Latzke,Liu_Bin,Absor4,AndorB,Cheng}, the spin splitting is identified on the $K$ and $Q$ points in the VBM and CBM, respectively. In the case of the polar WSSe ML, large spin splitting up to 0.45 eV is observed on the $K$ point in the VBM, which is comparable with that in the case of the non-polar WS$_{2}$ ML ($\Delta E_{K,\texttt{VBM}}=0.43$ eV). However, in the $Q$ point of the CBM, the spin splitting in the case of the polar WSSe ML ($\Delta E_{Q,\texttt{CBM}}=0.20$ eV) is smaller than that in the case of the non-polar  WS$_{2}$ ($\Delta E_{Q,\texttt{CBM}}=0.33$ eV). Interestingly, a sizable spin splitting around the $\Gamma$ point in the VBM that is not found in the case of the non-polar WS$_{2}$ ML is observed in the case of the polar WSSe ML [Fig. 2]. This splitting is referred as Rashba splitting, which is consistent with that previously reported by Cheng $at$ $al$. \cite{Cheng}. Because the energy level of the $K$ point is close to that of the $\Gamma$ point in the VBM, interplay between the Zeeman-like spin  splitting in the $K$ point and the Rashba spin splitting around the $\Gamma$ point is achieved, which is expected to play a significant role in the spintronics phenomena such as spin-conserving scattering.   

\subsection{Tunable the electronic and spin splitting properties by the strain}

Because the polarity plays a significant role in the electronic and spin splitting properties of the TMDs ML, controlling the polarity is expected to induce useful properties for spintronics. Here, strain is an effective method to tune the polarity, which is achieved by applying substrates \cite {Xenogiannopoulou,Aretouli,Brumme}. To this aim, we use a wide range of biaxial strains (up to 8\%) in the polar WSSe ML by tuning the planar lattice parameter. The range of the strain considered in this work was chosen because the broken of inter-atomic bonds may occur for the larger strain. For example, in the MoS$_{2}$ ML, the breaking of the interatomic bonds is achieved at an effective strain of 6 to 11 \% \cite{Bertolazzi}. Because the WSSe ML is polar material, the polar substrates having an intrinsic dipole moment such as wurtzite semiconductors \cite {Defo} is suitable for inducing the strain. By choosing AlN as an example of the polar substrates ($a_{0}$=3.11 \AA\ \cite{Nilsson}), a lattice constant mismatch of about 4.2 \% is achieved on the interface of the polar WSSe ML/AlN (0001), which is still within the range of the strain considered in the present study. We define the degree of in-plane biaxial strain as $\epsilon=(a-a_0)/a_0$, where $a_0$ is the unstrained in-plane lattice constant. Here, we studied the following two different cases: the tensile strain, which increases the in-plane lattice constant $a$, and compressive strain, which decreases $a$.

\begin{figure}
	\centering
		\includegraphics[width=1.0\textwidth]{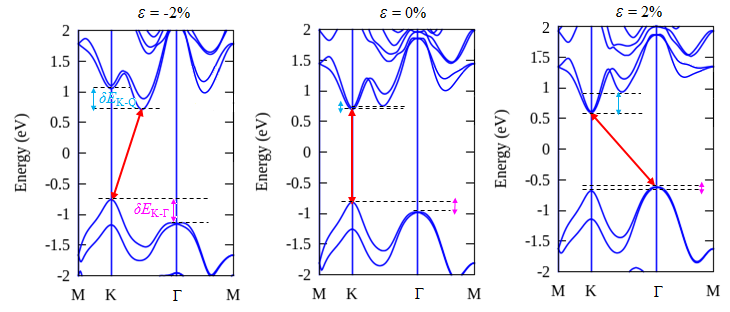}
	\caption{Electronic band structures of the strained polar WSSe ML with $\epsilon=-2\%$ (left), $\epsilon=0\%$ (center), and $\epsilon=2\%$ (right) are given. The red, pink, and blue arrows indicate the band gap, energy difference between the $K$ and $\Gamma$ points ($\delta E_{K-\Gamma}$), and energy difference between the $K$ and $Q$ points ($\delta E_{K-Q}$), respectively. The calculations are performed by including the spin-orbit coupling.}
	\label{figure:Figure3}
\end{figure}

Under the tensile strain, we find that the out-of-plane interlayer distance difference $\Delta d_{\bot}$ is decreased [Fig. 4(a)], and, consequently, the hybridization between the out-of-plane bonding states ( $d_{z^2}$, $p_{z}$ ) is reduced, while the hybridization between the in-plane bonding states [ $d_{{x^2}-{y^2}}+d_{xy}$, $S$ ${p_x}+{p_y}$ ] is strengthened. As a result, energy level of the $\Gamma$ point is higher than that of the $K$ point in the VBM [Fig. 3]. On the other hand, the compressive strain shifts the energy level of the $d_{{x^2}-{y^2}}+d_{xy}$ anti-bonding states to be lower than that of the $d_{z^2}$ anti-bonding states, resulting in that the energy level of the $Q$ point is lower than that of the $K$ point in the CBM. The shift of the energy level at the high symmetry points in the CBM and VBM by the strain has two important effects: (i) indirect band gap [Fig. 3], and (ii) the change of energy difference between the high symmetry points in the CBM and VBM [Fig. 4(b)]. We emphasized here that transition from the direct to the indirect band gap is achieved on a substantial critical strain, which is observed on -0.25 \% and 1.9 \% for the compressive and tensile strains, respectively. These values are slightly different from that seen on the non-polar WS$_{2}$ ML [(-1.2 \% and 2.3 \%) \cite{Absor4}], but consistent with previous results reported by Defo $at$. $al$. \cite{Defo}. Remarkably, tuning the polarity by the strain significantly modifies the electronic properties of the polar WSSe ML. 

\begin{figure}
	\centering
		\includegraphics[width=0.9\textwidth]{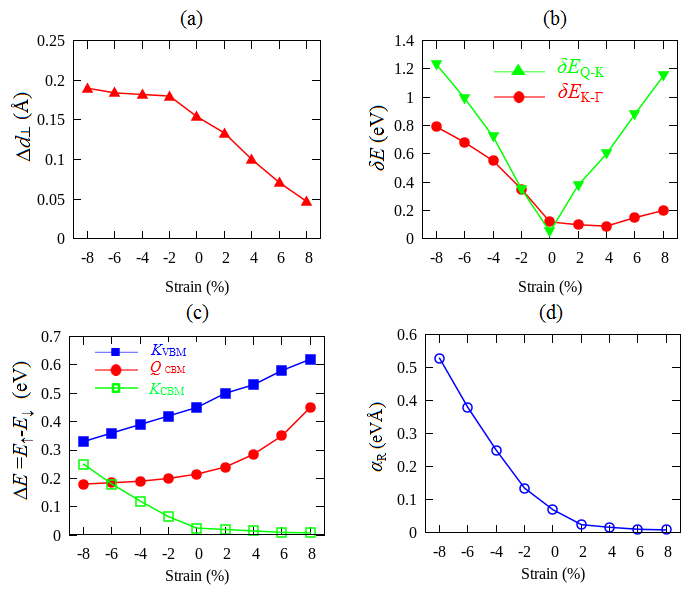}
	\caption{(a) Strain-dependent of the out-of-plane interlayer distance difference ($\Delta d_{\bot}$) in the polar WSSe ML is shown. (b) The calculated value of the energy difference between the $K$ and $\Gamma$ points ($\delta E_{K-\Gamma}$) in the VBM, between the $K$ and $Q$ points in the CBM ($\delta E_{K-Q}$), and between the $Q$ and $\Lambda$ points in the CBM ($\delta E_{Q-\Lambda}$) as a function of strain. (c) Strain-dependent of the spin splitting energy $\Delta E=\left|E_{\uparrow}-E_{\downarrow}\right|$ at the high symmetry points in the CBM and VBM. (d) The Rashba parameter ($\alpha_{R}$) as a function of strain, calculated around the $\Gamma$ point in the VBM.}
	\label{figure:Figure4}
\end{figure}

The strong modification of the electronic properties of the polar WSSe ML by the biaxial strain significantly changes the spin splitting properties of the bands. Due to the increased overlap of the in-plane bonding states [ $d_{{x^2}-{y^2}}+d_{xy}$, $S$ ${p_x}+{p_y}$ ] by the tensile strain, strong enhancement of the spin splitting is observed on the $K$ point in the VBM [Fig. 4(c)]. On the other hand, introducing the compressive strain increases the overlap of the out-of-plane bonding states ( $d_{z^2}$, $p_{z}$ ), which contributes to the increasing of the spin splitting around the $\Gamma$ point in the VBM [Fig. 2]. Consistently, the enhancement of the spin splitting is also identified in the CBM, which is observed on the $Q$ point under the tensile strain, and on the $K$ point under the compressive strain. These considerably changes of the spin splitting in the polar WSSe ML by applying the strain is expected to be useful for spintronics applications.

To better understand the nature of the observed spin splitting in the strained polar WSSe ML, we calculated the spin textures of the spin-split bands. Here, we focused on the spin textures in the VBM because of the enhanced spin splitting [Fig. 4(c)]. By considering the spin textures located on the 0.45 eV below the VBM, we find sixfold symmetry of spin-split hole pockets in the equilibrium as well as the strained systems. In the case of the equilibrium system, these hole pockets are clearly visible around the $K$ point, exhibiting fully out-of-plane spin polarization [Fig. 5(b)]. On the other hand, the spin-split hole pockets show in-plane polarization around the $\Gamma$ point, which is similar to the Rashba type spin textures\cite{Cheng}. Introducing strain subsequently modifies the spin textures due to the shifting in energy of the VBM. In the case of the compressive strain, the energy level of the $K$ point in the VBM is much higher than that of the $\Gamma$ point, thus the spin textures is dominated by the out-of-plane polarization of the spin-split hole pockets around the $K$ point [Fig. 5(a)]. On the contrary, in the case of the tensile strain, the energy level of the $\Gamma$ point is close to that of the $K$ point [Fig. 3], resulting in that the spin textures are characterized by the out-of-plane and in-plane polarizations of the spin-split hole pockets around the $K$ and $\Gamma$ points, respectively [Fig. 5(c)]. Because the energy difference between the $K$ and $\Gamma$ points is small [Fig. 4(b)], strong admixtures between the out-of-plane and in-plane spin polarized states are achieved, which is expected to influence the properties of spin relaxation and intervalley scattering times through the Dyakonov-Perel spin relaxation mechanism \cite{Schmidt,L_Yang} .
 
\begin{figure}
	\centering
		\includegraphics[width=1.0\textwidth]{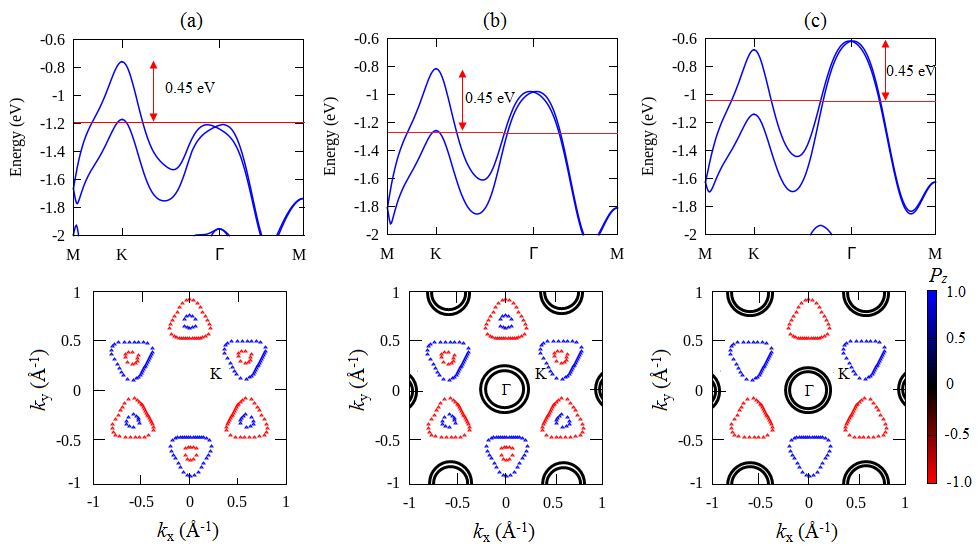}
	\caption{The spin textures of the strained polar WSSe ML evaluated on the VBM: (a) $\epsilon=-2\%$, (b) $\epsilon=0\%$, and (c) $\epsilon=2\%$. The spin textures are calculated on the constant energy located on 0.45 eV below the VBM. The position of $\Gamma$ and $K$ points are indicated. Color scale shows expectation value of spin in the out-of-plane direction ($P_{z}$). Here, red and blue colors indicate the fully out-of-plane orientations of the spin-up ($P_{z}=1$) and spin-down ($P_{z}=-1$) states, respectively, while the black colors indicate clock-wise and anti-clockwise in-plane orientations of the spin states ($P_{z}=0$).}
	\label{figure:Figure5}
\end{figure}
 
To clarify the origin of the observed spin splitting and the spin-polarized states in our calculational results, we consider our system based on the symmetry arguments. As mentioned before that the polar WSSe ML belongs to the  $C_{3v}$ symmetry group. Here, the symmetry itself consists of a $C_3$ rotation and a mirror symmetry operation $M_{y-z}:x\longrightarrow-x$, where $x$ is along the $\Gamma$-$K$ direction. In the case of spin $\frac{1}{2}$ electrons, $C_3$ and $M_{y-z}$ can be represented as $\exp^{-i\sigma_{z}\frac{\pi}{3}}$ and $i\sigma_{x}$, respectively, where $\sigma_{x,y,z}$ are the Pauli matrices for spin degree of freedom. Additionally, the anti-unitary time reversal operator $T$ that is represented by $i\sigma_{y}K$, where $K$ is complex conjugation, commuts with both $C_3$ and $M_{y-z}$. The $\vec{k}\cdot \vec{p}$ Hamiltonian $H(\vec{k})$ up to cubic $k$-terms can be constructed around the $\Gamma$ point by using invariant condition under $C_3$, $M_{y-z}$, and $T$ operations, which can be expressed as \cite {Fu,Yuan,Vajna}
\begin{equation}
\label{1}
H(\vec{k})=E_{0}(k)+\alpha_{R}(k_{x}\sigma_{y}-k_{y}\sigma_{x})+\beta_{k}(3{k^2_{x}}-{k^2_{y}})k_{y}\sigma_{z},
\end{equation}
where $E_{0}(k)={k^2}/2{m^*}$ and $k=\sqrt{{k^2_{x}}+{k^2_{y}}}$. Here, the second term in the $H(\vec{k})$ is the Rashba term characterized by Rashba parameter, $\alpha_{R}$, which induces in-plane component of the spin polarization. On the other hand, the third term in the $H(\vec{k})$ is the warping term characterized by warping parameter, $\beta_{k}$, which contributes to the out-of-plane component of spin polarization. It is noted here that the Rashba parameter $\alpha_{R}$ is induced by out-of-plane potential gradient asymmetry, while the warping parameter $\beta_{k}$ is mainly contributed from in-plane potential gradient asymmetry. Solving the eigenvalues problem for the Hamiltonian of Eqs. (\ref{1}) gives the following split energies:
\begin{equation}
\label{2}
E_{\pm}(k,\theta)=E_{0}(k)\pm\sqrt{{\alpha^{2}_{R}}{k^2}+{\beta^{2}_{k}{k^6}{\cos^{2}(3\theta)}}},
\end{equation}
where $\theta =\tan^{-1}(k_{y}/k_{x})$ is the azimuth angle of momentum $k$ with respect to the $x$ axis along the $\Gamma$-$K$ direction. The subscripts + and - denote states for the upper and lower bands, respectively.
The spin polarization vector $\vec{P}_{\pm}(k,\theta)$ can be evaluated using the averaged components of the spin operator $\left\langle \vec{\sigma}\right\rangle$, which turns out to be 
\begin{equation}
\label{3}
\vec{P}_{\pm}(k,\theta)= \pm[\alpha_{R}\sin\theta,-\alpha_{R}\cos\theta,-\beta_{k}\sin(3\theta)].
\end{equation} 

In the case of the non-polar WS${_2}$ ML, the mirror symmetry operation $M_{x-y}:z\longrightarrow-z$ in the $D_{3h}$ point group [Fig. 1(a)] suppresses the Rashba term in the $H(\vec{k})$, leading to the fact that only the third term in the Eq. (\ref{1}) remains. As a result, zero spin splitting is observed at $\theta=n\pi/3$, where $n$ is an integer number,  which is consistent with the observed spin degeneracy in the band structures along the $\Gamma$-$M$ direction shown in Fig. 2. However, the broken mirror symmetry $M_{x-y}$ in the case of the polar WSSe ML lifts the spin degeneracy along the $\Gamma$-$M$ [Fig. 2], which is due to the second term of the Eq. (\ref{1}). Furthermore, due to the last term of the Eq. (\ref{3}), fully-out-of plane spin polarization is visible at $\theta=(2n+1)\pi/6$, which is, in fact, consistent with our calculated results of the spin textures around the $K$ point in the VBM [Fig. (5)]. However, the in-plane spin polarization is observed in the spin-split bands around the $\Gamma$ point due to the first and second terms of Eq. (\ref{3}), which is consistent with our results shown in Fig. 5. Therefore, it can be concluded that the spin splitting and spin textures in our calculational results are consistent well with the simplified Hamiltonian.

\subsection{Discussion about possible application}

Before continuing our discussion on possible spintronic applications of the enhanced spin splitting in the strained polar WSSe ML, we briefly comment on the seemingly spin splitting features around the $\Gamma$ point in the VBM known as the Rashba splitting [see Fig. 2]. A comprehensive discussion of the Rashba splitting in the TMDs ML has been previously presented by Cheng $at$ $al$. \cite{Cheng}. However, they have not considered the effect of the strain, which motivated us to extend their study. Naturally, the non-zero of $\Delta d_{\bot}$ induces asymmetry of potential gradient perpendicular to the surface plane, leading to the strong hybridization between the out-of-plane bonding states [$d_{z^2}$, $p_z$] in the $\Gamma$ point. Consequently, the SOC leads to the Rashba spin splitting around the $\Gamma$ point. However, decreasing (increasing) $\Delta d_{\bot}$ by the compressive (tensile) strains [Fig. 4(a)], subsequently reduces (strengthens) the coupling between the out-of-plane bonding states, which decreases (increases) the Rashba splitting around the $\Gamma$ point [Fig. 3]. The considerably changes of the Rashba splitting by the strain is in fact consistent with modulation of the Rashba parameter $\alpha_{R}$ shown in Fig. 4(d). It is noted here that the calculated value of $\alpha_{R}$ in the present system is obtained by using the linear Rashba model throught relation defined by $\alpha_{R}=2E_{R}/k_{R}$, where $E_{R}$ and $k_{R}$ are the Rashba energy and momentum offset, respectively. Here, $E_{R}$ and $k_{R}$ can be directly evaluated from the band dispersion obtained by the first-principles calculation [Fig. 2]. By analyzing the spin-split states given in Fig. 4, we found that $\left|\beta_{k}/\alpha_{R}\right|\approx 1$ \% for small $k$, indicating that the contribution of $\beta_{k}$ to the spin splitting around the $\Gamma$ point is negligible, which confirms a consistency of our calculational results. Finally, the observed Rashba splitting in the present work also supports the recent prediction of the Rashba splitting on strained MoS$_{2}$/Bi(111) heterostructures \cite {Lee}, where the broken of the mirror symmetry by the substrates plays an important role in generating the Rashba splitting in the various TMDs ML. 

Here, we discuss the possible spintronic applications of the strained polar WSSe ML based on the features of the spin textures in the VBM. In the case of the equilibrium system, we found that the spin textures of the VBM are characterized by the out-of-plane and in-plane polarizations of the spin-split hole pockets around the $K$ and $\Gamma$ points, respectively [Fig. 5(b)]. These features of the spin textures are expected to induce spin-orbit field in the out-of-plane $B_{SO_{\bot}}$ and in-plane $B_{SO_{\|}}$ directions, respectively. The same orientation of the spin-orbit fields is also observed in the case of the tensile strain, which is due to the same features of the spin textures. Because the energy difference between the $K$ and $\Gamma$ points, $\delta E_{K-\Gamma}$, is small [Fig. 4(b)], strong admixtures between the out-of-plane and in-plane spin polarized states are achieved, leading to the fact that the coupling between $B_{SO_{\bot}}$ and $B_{SO_{\|}}$ around the $K$ and $\Gamma$ points, respectively, is strengthened. However, due to the substantially small spin splitting around the $\Gamma$ point, weak in-plane spin-orbit field $B_{SO_{\|}}$ is generated, inducing small misalignment of the total spin-orbit fields from the out-of-plane direction. Accordingly, Dyakonov-Perel spin relaxation mechanism implies that the spin relaxation times are much longer than that intervalley scattering times. This situation is qualitatively similar to the recent observation of the spin relaxation on dual gated exfoliated MoS${_2}$ ML reported by Schmidt $at$ $al$. \cite{Schmidt}, which found that applying ionic gating breaks the mirror symmetry to induces the enhanced spin relaxation. 

Interestingly, the spin textures of the VBM in the case of the compressive strain is dominated by the fully out-of-plane  polarization of the spin-split hole pockets around the $K$ point [Fig. 5(a)]. Here, the fully out-of-plane orientation of the spin-orbit fields $B_{SO_{\bot}}$ is generated, implying that unusually long spin relaxation times without intervalley scattering is achieved. This is supported by the fact that a similar mechanism behind the long spin relaxation times induced by out-of-plane spin polarization has been reported on the various TMDs ML \cite {Zhu,Schmidt,L_Yang,Liu,Absor4}, suggesting that the present system is promising for energy saving spintronics. 

It is pointed out here that our proposed approach for generating and modulating the spin splitting and spin-polarized states is not limited to the polar WSSe ML, but it can be generalized to a variety of SOC systems with the polar $MXY$ structures exhibiting the polarity-induced mirror symmetry breaking (for example, other the polar TMDS ML including WSTe, MoSSe and MoSTe whose electronic structure properties are similar to WSSe ML \cite{Cheng,Defo}). Together, these features seem to be promising for inducing new electronic properties which are useful for spintronic applications.

\section{CONCLUSION}
To summarize, we have investigated the effect of the polarity and its strain dependent on the electronic properties of the TMDs ML by using first-principles density functional theory calculations. We found that in addition to the established spin splitting along the $\Gamma$-$K$ line with fully-out-of-plane spin polarization, the presence of the mirror symmetry breaking in the polar TMDs ML leads to sizable spin splitting along the $\Gamma$-$M$ line exhibiting in-plane spin polarization. We also find that these splittings are effectively controlled by tuning the polarity realized by introducing biaxial strain. We clarified the origin of the spin splitting and spin polarization in our calculational results by using symmetry arguments combined with orbital hybridization analyses. The enhanced and sizable spin splittings found in the present study suggested that the strained polar TMDs ML systems are suitable for spintronic applications. Finally, we have identified the possible admixtures of the out-of-plane and in-plane spin-polarized states in the strained polar systems and discussed their implications to the spin relaxation involving intervalley scattering process through the Dyakonov-Perel mechanism. Our study clarified that the polarity-induced mirror symmetry breaking is an important parameter to control the spin splitting and spin relaxation in the TMDs ML, which is useful for designing future spintronic devices.

\begin{acknowledgments}

This work was supported by the Fundamental Reserach Grant (2017) funded by the ministry of research and technolgy and higher eduacation, Republic of Indonesia. The computations in this research were performed using the high performance computing facilities (DSDI) at Universitas Gadjah Mada, Indonesia. 

\end{acknowledgments}

\bibliography{Reference1}

\begin{thebibliography}{47}%
\makeatletter
\providecommand \@ifxundefined [1]{%
 \@ifx{#1\undefined}
}%
\providecommand \@ifnum [1]{%
 \ifnum #1\expandafter \@firstoftwo
 \else \expandafter \@secondoftwo
 \fi
}%
\providecommand \@ifx [1]{%
 \ifx #1\expandafter \@firstoftwo
 \else \expandafter \@secondoftwo
 \fi
}%
\providecommand \natexlab [1]{#1}%
\providecommand \enquote  [1]{``#1''}%
\providecommand \bibnamefont  [1]{#1}%
\providecommand \bibfnamefont [1]{#1}%
\providecommand \citenamefont [1]{#1}%
\providecommand \href@noop [0]{\@secondoftwo}%
\providecommand \href [0]{\begingroup \@sanitize@url \@href}%
\providecommand \@href[1]{\@@startlink{#1}\@@href}%
\providecommand \@@href[1]{\endgroup#1\@@endlink}%
\providecommand \@sanitize@url [0]{\catcode `\\12\catcode `\$12\catcode
  `\&12\catcode `\#12\catcode `\^12\catcode `\_12\catcode `\%12\relax}%
\providecommand \@@startlink[1]{}%
\providecommand \@@endlink[0]{}%
\providecommand \url  [0]{\begingroup\@sanitize@url \@url }%
\providecommand \@url [1]{\endgroup\@href {#1}{\urlprefix }}%
\providecommand \urlprefix  [0]{URL }%
\providecommand \Eprint [0]{\href }%
\providecommand \doibase [0]{http://dx.doi.org/}%
\providecommand \selectlanguage [0]{\@gobble}%
\providecommand \bibinfo  [0]{\@secondoftwo}%
\providecommand \bibfield  [0]{\@secondoftwo}%
\providecommand \translation [1]{[#1]}%
\providecommand \BibitemOpen [0]{}%
\providecommand \bibitemStop [0]{}%
\providecommand \bibitemNoStop [0]{.\EOS\space}%
\providecommand \EOS [0]{\spacefactor3000\relax}%
\providecommand \BibitemShut  [1]{\csname bibitem#1\endcsname}%
\let\auto@bib@innerbib\@empty
\bibitem [{\citenamefont {Kato}\ \emph {et~al.}(2004)\citenamefont {Kato},
  \citenamefont {Myers}, \citenamefont {Gossard},\ and\ \citenamefont
  {Awschalom}}]{Kato}%
  \BibitemOpen
  \bibfield  {author} {\bibinfo {author} {\bibfnamefont {Y.}~\bibnamefont
  {Kato}}, \bibinfo {author} {\bibfnamefont {R.~C.}\ \bibnamefont {Myers}},
  \bibinfo {author} {\bibfnamefont {A.~C.}\ \bibnamefont {Gossard}}, \ and\
  \bibinfo {author} {\bibfnamefont {D.~D.}\ \bibnamefont {Awschalom}},\
  }\bibfield  {title} {\enquote {\bibinfo {title} {Coherent spin manipulation
  without magnetic fields in strained semiconductors},}\ }\href {\doibase
  10.1038/nature02202} {\bibfield  {journal} {\bibinfo  {journal} {Nature}\
  }\textbf {\bibinfo {volume} {427}},\ \bibinfo {pages} {50} (\bibinfo {year}
  {2004})}\BibitemShut {NoStop}%
\bibitem [{\citenamefont {Kuhlen}\ \emph {et~al.}(2012)\citenamefont {Kuhlen},
  \citenamefont {Schmalbuch}, \citenamefont {Hagedorn}, \citenamefont
  {Schlammes}, \citenamefont {Patt}, \citenamefont {Lepsa}, \citenamefont
  {G\"untherodt},\ and\ \citenamefont {Beschoten}}]{Kuhlen}%
  \BibitemOpen
  \bibfield  {author} {\bibinfo {author} {\bibfnamefont {S.}~\bibnamefont
  {Kuhlen}}, \bibinfo {author} {\bibfnamefont {K.}~\bibnamefont {Schmalbuch}},
  \bibinfo {author} {\bibfnamefont {M.}~\bibnamefont {Hagedorn}}, \bibinfo
  {author} {\bibfnamefont {P.}~\bibnamefont {Schlammes}}, \bibinfo {author}
  {\bibfnamefont {M.}~\bibnamefont {Patt}}, \bibinfo {author} {\bibfnamefont
  {M.}~\bibnamefont {Lepsa}}, \bibinfo {author} {\bibfnamefont
  {G.}~\bibnamefont {G\"untherodt}}, \ and\ \bibinfo {author} {\bibfnamefont
  {B.}~\bibnamefont {Beschoten}},\ }\bibfield  {title} {\enquote {\bibinfo
  {title} {Electric field-driven coherent spin reorientation of optically
  generated electron spin packets in ingaas},}\ }\href {\doibase
  10.1103/PhysRevLett.109.146603} {\bibfield  {journal} {\bibinfo  {journal}
  {Phys. Rev. Lett.}\ }\textbf {\bibinfo {volume} {109}},\ \bibinfo {pages}
  {146603} (\bibinfo {year} {2012})}\BibitemShut {NoStop}%
\bibitem [{\citenamefont {Qi}, \citenamefont {Wu},\ and\ \citenamefont
  {Zhang}(2006)}]{Qi}%
  \BibitemOpen
  \bibfield  {author} {\bibinfo {author} {\bibfnamefont {X.-L.}\ \bibnamefont
  {Qi}}, \bibinfo {author} {\bibfnamefont {Y.-S.}\ \bibnamefont {Wu}}, \ and\
  \bibinfo {author} {\bibfnamefont {S.-C.}\ \bibnamefont {Zhang}},\ }\bibfield
  {title} {\enquote {\bibinfo {title} {Topological quantization of the spin
  hall effect in two-dimensional paramagnetic semiconductors},}\ }\href
  {\doibase 10.1103/PhysRevB.74.085308} {\bibfield  {journal} {\bibinfo
  {journal} {Phys. Rev. B}\ }\textbf {\bibinfo {volume} {74}},\ \bibinfo
  {pages} {085308} (\bibinfo {year} {2006})}\BibitemShut {NoStop}%
\bibitem [{\citenamefont {Datta}\ and\ \citenamefont {Das}(1990)}]{Datta}%
  \BibitemOpen
  \bibfield  {author} {\bibinfo {author} {\bibfnamefont {S.}~\bibnamefont
  {Datta}}\ and\ \bibinfo {author} {\bibfnamefont {B.}~\bibnamefont {Das}},\
  }\bibfield  {title} {\enquote {\bibinfo {title} {Electronic analog of the
  electro‐optic modulator},}\ }\href {\doibase
  http://dx.doi.org/10.1063/1.102730} {\bibfield  {journal} {\bibinfo
  {journal} {Appl. Phys. Lett.}\ }\textbf {\bibinfo {volume} {56}},\ \bibinfo
  {pages} {665--667} (\bibinfo {year} {1990})}\BibitemShut {NoStop}%
\bibitem [{\citenamefont {Nitta}\ \emph {et~al.}(1997)\citenamefont {Nitta},
  \citenamefont {Akazaki}, \citenamefont {Takayanagi},\ and\ \citenamefont
  {Enoki}}]{NittaA}%
  \BibitemOpen
  \bibfield  {author} {\bibinfo {author} {\bibfnamefont {J.}~\bibnamefont
  {Nitta}}, \bibinfo {author} {\bibfnamefont {T.}~\bibnamefont {Akazaki}},
  \bibinfo {author} {\bibfnamefont {H.}~\bibnamefont {Takayanagi}}, \ and\
  \bibinfo {author} {\bibfnamefont {T.}~\bibnamefont {Enoki}},\ }\bibfield
  {title} {\enquote {\bibinfo {title} {Gate control of spin-orbit interaction
  in an inverted
  i${\mathrm{n}}_{0.53}$g${\mathrm{a}}_{0.47}$as/i${\mathrm{n}}_{0.52}$a${\mathrm{l}}_{0.48}$as
  heterostructure},}\ }\href {\doibase 10.1103/PhysRevLett.78.1335} {\bibfield
  {journal} {\bibinfo  {journal} {Phys. Rev. Lett.}\ }\textbf {\bibinfo
  {volume} {78}},\ \bibinfo {pages} {1335--1338} (\bibinfo {year}
  {1997})}\BibitemShut {NoStop}%
\bibitem [{\citenamefont {Novoselov}\ \emph {et~al.}(2004)\citenamefont
  {Novoselov}, \citenamefont {Geim}, \citenamefont {Morozov}, \citenamefont
  {Jiang}, \citenamefont {Zhang}, \citenamefont {Dubonos}, \citenamefont
  {Grigorieva},\ and\ \citenamefont {Firsov}}]{Novoselov}%
  \BibitemOpen
  \bibfield  {author} {\bibinfo {author} {\bibfnamefont {K.~S.}\ \bibnamefont
  {Novoselov}}, \bibinfo {author} {\bibfnamefont {A.~K.}\ \bibnamefont {Geim}},
  \bibinfo {author} {\bibfnamefont {S.~V.}\ \bibnamefont {Morozov}}, \bibinfo
  {author} {\bibfnamefont {D.}~\bibnamefont {Jiang}}, \bibinfo {author}
  {\bibfnamefont {Y.}~\bibnamefont {Zhang}}, \bibinfo {author} {\bibfnamefont
  {S.~V.}\ \bibnamefont {Dubonos}}, \bibinfo {author} {\bibfnamefont {I.~V.}\
  \bibnamefont {Grigorieva}}, \ and\ \bibinfo {author} {\bibfnamefont {A.~A.}\
  \bibnamefont {Firsov}},\ }\bibfield  {title} {\enquote {\bibinfo {title}
  {Electric field effect in atomically thin carbon films},}\ }\href {\doibase
  10.1126/science.1102896} {\bibfield  {journal} {\bibinfo  {journal}
  {Science}\ }\textbf {\bibinfo {volume} {306}},\ \bibinfo {pages} {666}
  (\bibinfo {year} {2004})}\BibitemShut {NoStop}%
\bibitem [{\citenamefont {Cahangirov}\ \emph {et~al.}(2009)\citenamefont
  {Cahangirov}, \citenamefont {Topsakal}, \citenamefont {Akt\"urk},
  \citenamefont {\ifmmode~\mbox{\c{S}}\else \c{S}\fi{}ahin},\ and\
  \citenamefont {Ciraci}}]{Cahangirov}%
  \BibitemOpen
  \bibfield  {author} {\bibinfo {author} {\bibfnamefont {S.}~\bibnamefont
  {Cahangirov}}, \bibinfo {author} {\bibfnamefont {M.}~\bibnamefont
  {Topsakal}}, \bibinfo {author} {\bibfnamefont {E.}~\bibnamefont {Akt\"urk}},
  \bibinfo {author} {\bibfnamefont {H.}~\bibnamefont
  {\ifmmode~\mbox{\c{S}}\else \c{S}\fi{}ahin}}, \ and\ \bibinfo {author}
  {\bibfnamefont {S.}~\bibnamefont {Ciraci}},\ }\bibfield  {title} {\enquote
  {\bibinfo {title} {Two- and one-dimensional honeycomb structures of silicon
  and germanium},}\ }\href {\doibase 10.1103/PhysRevLett.102.236804} {\bibfield
   {journal} {\bibinfo  {journal} {Phys. Rev. Lett.}\ }\textbf {\bibinfo
  {volume} {102}},\ \bibinfo {pages} {236804} (\bibinfo {year}
  {2009})}\BibitemShut {NoStop}%
\bibitem [{\citenamefont {Min}\ \emph {et~al.}(2006)\citenamefont {Min},
  \citenamefont {Hill}, \citenamefont {Sinitsyn}, \citenamefont {Sahu},
  \citenamefont {Kleinman},\ and\ \citenamefont {MacDonald}}]{Min}%
  \BibitemOpen
  \bibfield  {author} {\bibinfo {author} {\bibfnamefont {H.}~\bibnamefont
  {Min}}, \bibinfo {author} {\bibfnamefont {J.~E.}\ \bibnamefont {Hill}},
  \bibinfo {author} {\bibfnamefont {N.~A.}\ \bibnamefont {Sinitsyn}}, \bibinfo
  {author} {\bibfnamefont {B.~R.}\ \bibnamefont {Sahu}}, \bibinfo {author}
  {\bibfnamefont {L.}~\bibnamefont {Kleinman}}, \ and\ \bibinfo {author}
  {\bibfnamefont {A.~H.}\ \bibnamefont {MacDonald}},\ }\bibfield  {title}
  {\enquote {\bibinfo {title} {Intrinsic and rashba spin-orbit interactions in
  graphene sheets},}\ }\href {\doibase 10.1103/PhysRevB.74.165310} {\bibfield
  {journal} {\bibinfo  {journal} {Phys. Rev. B}\ }\textbf {\bibinfo {volume}
  {74}},\ \bibinfo {pages} {165310} (\bibinfo {year} {2006})}\BibitemShut
  {NoStop}%
\bibitem [{\citenamefont {Liu}, \citenamefont {Feng},\ and\ \citenamefont
  {Yao}(2011)}]{LiuC}%
  \BibitemOpen
  \bibfield  {author} {\bibinfo {author} {\bibfnamefont {C.-C.}\ \bibnamefont
  {Liu}}, \bibinfo {author} {\bibfnamefont {W.}~\bibnamefont {Feng}}, \ and\
  \bibinfo {author} {\bibfnamefont {Y.}~\bibnamefont {Yao}},\ }\bibfield
  {title} {\enquote {\bibinfo {title} {Quantum spin hall effect in silicene and
  two-dimensional germanium},}\ }\href {\doibase
  10.1103/PhysRevLett.107.076802} {\bibfield  {journal} {\bibinfo  {journal}
  {Phys. Rev. Lett.}\ }\textbf {\bibinfo {volume} {107}},\ \bibinfo {pages}
  {076802} (\bibinfo {year} {2011})}\BibitemShut {NoStop}%
\bibitem [{\citenamefont {Kane}\ and\ \citenamefont {Mele}(2005)}]{Kane}%
  \BibitemOpen
  \bibfield  {author} {\bibinfo {author} {\bibfnamefont {C.~L.}\ \bibnamefont
  {Kane}}\ and\ \bibinfo {author} {\bibfnamefont {E.~J.}\ \bibnamefont
  {Mele}},\ }\bibfield  {title} {\enquote {\bibinfo {title} {Quantum spin hall
  effect in graphene},}\ }\href {\doibase 10.1103/PhysRevLett.95.226801}
  {\bibfield  {journal} {\bibinfo  {journal} {Phys. Rev. Lett.}\ }\textbf
  {\bibinfo {volume} {95}},\ \bibinfo {pages} {226801} (\bibinfo {year}
  {2005})}\BibitemShut {NoStop}%
\bibitem [{\citenamefont {Xiao}\ \emph {et~al.}(2012)\citenamefont {Xiao},
  \citenamefont {Liu}, \citenamefont {Feng}, \citenamefont {Xu},\ and\
  \citenamefont {Yao}}]{Xiao}%
  \BibitemOpen
  \bibfield  {author} {\bibinfo {author} {\bibfnamefont {D.}~\bibnamefont
  {Xiao}}, \bibinfo {author} {\bibfnamefont {G.-B.}\ \bibnamefont {Liu}},
  \bibinfo {author} {\bibfnamefont {W.}~\bibnamefont {Feng}}, \bibinfo {author}
  {\bibfnamefont {X.}~\bibnamefont {Xu}}, \ and\ \bibinfo {author}
  {\bibfnamefont {W.}~\bibnamefont {Yao}},\ }\bibfield  {title} {\enquote
  {\bibinfo {title} {Coupled spin and valley physics in monolayers of
  ${\mathrm{mos}}_{2}$ and other group-vi dichalcogenides},}\ }\href {\doibase
  10.1103/PhysRevLett.108.196802} {\bibfield  {journal} {\bibinfo  {journal}
  {Phys. Rev. Lett.}\ }\textbf {\bibinfo {volume} {108}},\ \bibinfo {pages}
  {196802} (\bibinfo {year} {2012})}\BibitemShut {NoStop}%
\bibitem [{\citenamefont {Xu}\ \emph {et~al.}(2014)\citenamefont {Xu},
  \citenamefont {Yao}, \citenamefont {Xiao},\ and\ \citenamefont {Heinz}}]{Xu}%
  \BibitemOpen
  \bibfield  {author} {\bibinfo {author} {\bibfnamefont {X.~D.}\ \bibnamefont
  {Xu}}, \bibinfo {author} {\bibfnamefont {W.}~\bibnamefont {Yao}}, \bibinfo
  {author} {\bibfnamefont {D.}~\bibnamefont {Xiao}}, \ and\ \bibinfo {author}
  {\bibfnamefont {T.~F.}\ \bibnamefont {Heinz}},\ }\bibfield  {title} {\enquote
  {\bibinfo {title} {Spin and pseudospins in layered transition metal
  dichalcogenides},}\ }\href {\doibase 10.1038/nphys2942} {\bibfield  {journal}
  {\bibinfo  {journal} {Nat. Phyiscs}\ }\textbf {\bibinfo {volume} {10}},\
  \bibinfo {pages} {343} (\bibinfo {year} {2014})}\BibitemShut {NoStop}%
\bibitem [{\citenamefont {Yuan}\ \emph {et~al.}(2013)\citenamefont {Yuan},
  \citenamefont {Bahramy}, \citenamefont {Morimoto}, \citenamefont {Wu},
  \citenamefont {Nomura}, \citenamefont {Yang}, \citenamefont {Shimotani},
  \citenamefont {Suzuki}, \citenamefont {Toh}, \citenamefont {Kloc},
  \citenamefont {Xu}, \citenamefont {Arita}, \citenamefont {Nagaosa},\ and\
  \citenamefont {Iwasa}}]{Yuan}%
  \BibitemOpen
  \bibfield  {author} {\bibinfo {author} {\bibfnamefont {H.}~\bibnamefont
  {Yuan}}, \bibinfo {author} {\bibfnamefont {M.~S.}\ \bibnamefont {Bahramy}},
  \bibinfo {author} {\bibfnamefont {K.}~\bibnamefont {Morimoto}}, \bibinfo
  {author} {\bibfnamefont {S.}~\bibnamefont {Wu}}, \bibinfo {author}
  {\bibfnamefont {K.}~\bibnamefont {Nomura}}, \bibinfo {author} {\bibfnamefont
  {B.-J.}\ \bibnamefont {Yang}}, \bibinfo {author} {\bibfnamefont
  {H.}~\bibnamefont {Shimotani}}, \bibinfo {author} {\bibfnamefont
  {R.}~\bibnamefont {Suzuki}}, \bibinfo {author} {\bibfnamefont
  {M.}~\bibnamefont {Toh}}, \bibinfo {author} {\bibfnamefont {C.}~\bibnamefont
  {Kloc}}, \bibinfo {author} {\bibfnamefont {X.}~\bibnamefont {Xu}}, \bibinfo
  {author} {\bibfnamefont {R.}~\bibnamefont {Arita}}, \bibinfo {author}
  {\bibfnamefont {N.}~\bibnamefont {Nagaosa}}, \ and\ \bibinfo {author}
  {\bibfnamefont {Y.}~\bibnamefont {Iwasa}},\ }\bibfield  {title} {\enquote
  {\bibinfo {title} {Zeeman-type spin splitting controlled by an electric
  field},}\ }\href {\doibase 10.1038/nphys2691} {\bibfield  {journal} {\bibinfo
   {journal} {Nat. Phyiscs}\ }\textbf {\bibinfo {volume} {9}},\ \bibinfo
  {pages} {563} (\bibinfo {year} {2013})}\BibitemShut {NoStop}%
\bibitem [{\citenamefont {Zhu}, \citenamefont {Cheng},\ and\ \citenamefont
  {Schwingenschl\"ogl}(2011)}]{Zhu}%
  \BibitemOpen
  \bibfield  {author} {\bibinfo {author} {\bibfnamefont {Z.~Y.}\ \bibnamefont
  {Zhu}}, \bibinfo {author} {\bibfnamefont {Y.~C.}\ \bibnamefont {Cheng}}, \
  and\ \bibinfo {author} {\bibfnamefont {U.}~\bibnamefont
  {Schwingenschl\"ogl}},\ }\bibfield  {title} {\enquote {\bibinfo {title}
  {Giant spin-orbit-induced spin splitting in two-dimensional transition-metal
  dichalcogenide semiconductors},}\ }\href {\doibase
  10.1103/PhysRevB.84.153402} {\bibfield  {journal} {\bibinfo  {journal} {Phys.
  Rev. B}\ }\textbf {\bibinfo {volume} {84}},\ \bibinfo {pages} {153402}
  (\bibinfo {year} {2011})}\BibitemShut {NoStop}%
\bibitem [{\citenamefont {Latzke}\ \emph {et~al.}(2015)\citenamefont {Latzke},
  \citenamefont {Zhang}, \citenamefont {Suslu}, \citenamefont {Chang},
  \citenamefont {Lin}, \citenamefont {Jeng}, \citenamefont {Tongay},
  \citenamefont {Wu}, \citenamefont {Bansil},\ and\ \citenamefont
  {Lanzara}}]{Latzke}%
  \BibitemOpen
  \bibfield  {author} {\bibinfo {author} {\bibfnamefont {D.~W.}\ \bibnamefont
  {Latzke}}, \bibinfo {author} {\bibfnamefont {W.}~\bibnamefont {Zhang}},
  \bibinfo {author} {\bibfnamefont {A.}~\bibnamefont {Suslu}}, \bibinfo
  {author} {\bibfnamefont {T.-R.}\ \bibnamefont {Chang}}, \bibinfo {author}
  {\bibfnamefont {H.}~\bibnamefont {Lin}}, \bibinfo {author} {\bibfnamefont
  {H.-T.}\ \bibnamefont {Jeng}}, \bibinfo {author} {\bibfnamefont
  {S.}~\bibnamefont {Tongay}}, \bibinfo {author} {\bibfnamefont
  {J.}~\bibnamefont {Wu}}, \bibinfo {author} {\bibfnamefont {A.}~\bibnamefont
  {Bansil}}, \ and\ \bibinfo {author} {\bibfnamefont {A.}~\bibnamefont
  {Lanzara}},\ }\bibfield  {title} {\enquote {\bibinfo {title} {Electronic
  structure, spin-orbit coupling, and interlayer interaction in bulk
  ${\mathrm{mos}}_{2}$ and ${\mathrm{ws}}_{2}$},}\ }\href {\doibase
  10.1103/PhysRevB.91.235202} {\bibfield  {journal} {\bibinfo  {journal} {Phys.
  Rev. B}\ }\textbf {\bibinfo {volume} {91}},\ \bibinfo {pages} {235202}
  (\bibinfo {year} {2015})}\BibitemShut {NoStop}%
\bibitem [{\citenamefont {Liu}\ \emph {et~al.}(2013)\citenamefont {Liu},
  \citenamefont {Shan}, \citenamefont {Yao}, \citenamefont {Yao},\ and\
  \citenamefont {Xiao}}]{Liu_Bin}%
  \BibitemOpen
  \bibfield  {author} {\bibinfo {author} {\bibfnamefont {G.-B.}\ \bibnamefont
  {Liu}}, \bibinfo {author} {\bibfnamefont {W.-Y.}\ \bibnamefont {Shan}},
  \bibinfo {author} {\bibfnamefont {Y.}~\bibnamefont {Yao}}, \bibinfo {author}
  {\bibfnamefont {W.}~\bibnamefont {Yao}}, \ and\ \bibinfo {author}
  {\bibfnamefont {D.}~\bibnamefont {Xiao}},\ }\bibfield  {title} {\enquote
  {\bibinfo {title} {Three-band tight-binding model for monolayers of group-vib
  transition metal dichalcogenides},}\ }\href {\doibase
  10.1103/PhysRevB.88.085433} {\bibfield  {journal} {\bibinfo  {journal} {Phys.
  Rev. B}\ }\textbf {\bibinfo {volume} {88}},\ \bibinfo {pages} {085433}
  (\bibinfo {year} {2013})}\BibitemShut {NoStop}%
\bibitem [{\citenamefont {Absor}\ \emph
  {et~al.}(2016{\natexlab{a}})\citenamefont {Absor}, \citenamefont {Kotaka},
  \citenamefont {Ishii},\ and\ \citenamefont {Saito}}]{Absor4}%
  \BibitemOpen
  \bibfield  {author} {\bibinfo {author} {\bibfnamefont {M.~A.~U.}\
  \bibnamefont {Absor}}, \bibinfo {author} {\bibfnamefont {H.}~\bibnamefont
  {Kotaka}}, \bibinfo {author} {\bibfnamefont {F.}~\bibnamefont {Ishii}}, \
  and\ \bibinfo {author} {\bibfnamefont {M.}~\bibnamefont {Saito}},\ }\bibfield
   {title} {\enquote {\bibinfo {title} {Strain-controlled spin splitting in the
  conduction band of monolayer ${\text{ws}}_{2}$},}\ }\href {\doibase
  10.1103/PhysRevB.94.115131} {\bibfield  {journal} {\bibinfo  {journal} {Phys.
  Rev. B}\ }\textbf {\bibinfo {volume} {94}},\ \bibinfo {pages} {115131}
  (\bibinfo {year} {2016}{\natexlab{a}})}\BibitemShut {NoStop}%
\bibitem [{\citenamefont {Chu}\ \emph {et~al.}(2014)\citenamefont {Chu},
  \citenamefont {Li}, \citenamefont {Wu}, \citenamefont {Niu}, \citenamefont
  {Yao}, \citenamefont {Xu},\ and\ \citenamefont {Zhang}}]{Chu}%
  \BibitemOpen
  \bibfield  {author} {\bibinfo {author} {\bibfnamefont {R.-L.}\ \bibnamefont
  {Chu}}, \bibinfo {author} {\bibfnamefont {X.}~\bibnamefont {Li}}, \bibinfo
  {author} {\bibfnamefont {S.}~\bibnamefont {Wu}}, \bibinfo {author}
  {\bibfnamefont {Q.}~\bibnamefont {Niu}}, \bibinfo {author} {\bibfnamefont
  {W.}~\bibnamefont {Yao}}, \bibinfo {author} {\bibfnamefont {X.}~\bibnamefont
  {Xu}}, \ and\ \bibinfo {author} {\bibfnamefont {C.}~\bibnamefont {Zhang}},\
  }\bibfield  {title} {\enquote {\bibinfo {title} {Valley-splitting and
  valley-dependent inter-landau-level optical transitions in monolayer
  ${\mathrm{mos}}_{2}$ quantum hall systems},}\ }\href {\doibase
  10.1103/PhysRevB.90.045427} {\bibfield  {journal} {\bibinfo  {journal} {Phys.
  Rev. B}\ }\textbf {\bibinfo {volume} {90}},\ \bibinfo {pages} {045427}
  (\bibinfo {year} {2014})}\BibitemShut {NoStop}%
\bibitem [{\citenamefont {Bromley}, \citenamefont {Murray},\ and\ \citenamefont
  {Yoffe}(1972)}]{Bromley}%
  \BibitemOpen
  \bibfield  {author} {\bibinfo {author} {\bibfnamefont {R.~A.}\ \bibnamefont
  {Bromley}}, \bibinfo {author} {\bibfnamefont {R.~B.}\ \bibnamefont {Murray}},
  \ and\ \bibinfo {author} {\bibfnamefont {A.~D.}\ \bibnamefont {Yoffe}},\
  }\bibfield  {title} {\enquote {\bibinfo {title} {The band structures of some
  transition metal dichalcogenides. iii. group via: trigonal prism
  materials},}\ }\href {http://stacks.iop.org/0022-3719/5/i=7/a=007} {\bibfield
   {journal} {\bibinfo  {journal} {Journal of Physics C: Solid State Physics}\
  }\textbf {\bibinfo {volume} {5}},\ \bibinfo {pages} {759} (\bibinfo {year}
  {1972})}\BibitemShut {NoStop}%
\bibitem [{\citenamefont {Ko\ifmmode~\acute{s}\else \'{s}\fi{}mider},
  \citenamefont {Gonz\'alez},\ and\ \citenamefont
  {Fern\'andez-Rossier}(2013)}]{Kosminder}%
  \BibitemOpen
  \bibfield  {author} {\bibinfo {author} {\bibfnamefont {K.}~\bibnamefont
  {Ko\ifmmode~\acute{s}\else \'{s}\fi{}mider}}, \bibinfo {author}
  {\bibfnamefont {J.~W.}\ \bibnamefont {Gonz\'alez}}, \ and\ \bibinfo {author}
  {\bibfnamefont {J.}~\bibnamefont {Fern\'andez-Rossier}},\ }\bibfield  {title}
  {\enquote {\bibinfo {title} {Large spin splitting in the conduction band of
  transition metal dichalcogenide monolayers},}\ }\href {\doibase
  10.1103/PhysRevB.88.245436} {\bibfield  {journal} {\bibinfo  {journal} {Phys.
  Rev. B}\ }\textbf {\bibinfo {volume} {88}},\ \bibinfo {pages} {245436}
  (\bibinfo {year} {2013})}\BibitemShut {NoStop}%
\bibitem [{\citenamefont {Cazalilla}, \citenamefont {Ochoa},\ and\
  \citenamefont {Guinea}(2014)}]{Cazalilla}%
  \BibitemOpen
  \bibfield  {author} {\bibinfo {author} {\bibfnamefont {M.~A.}\ \bibnamefont
  {Cazalilla}}, \bibinfo {author} {\bibfnamefont {H.}~\bibnamefont {Ochoa}}, \
  and\ \bibinfo {author} {\bibfnamefont {F.}~\bibnamefont {Guinea}},\
  }\bibfield  {title} {\enquote {\bibinfo {title} {Quantum spin hall effect in
  two-dimensional crystals of transition-metal dichalcogenides},}\ }\href
  {\doibase 10.1103/PhysRevLett.113.077201} {\bibfield  {journal} {\bibinfo
  {journal} {Phys. Rev. Lett.}\ }\textbf {\bibinfo {volume} {113}},\ \bibinfo
  {pages} {077201} (\bibinfo {year} {2014})}\BibitemShut {NoStop}%
\bibitem [{\citenamefont {Ma}\ \emph {et~al.}(2015)\citenamefont {Ma},
  \citenamefont {Kou}, \citenamefont {Li}, \citenamefont {Dai}, \citenamefont
  {Smith},\ and\ \citenamefont {Heine}}]{Ma}%
  \BibitemOpen
  \bibfield  {author} {\bibinfo {author} {\bibfnamefont {Y.}~\bibnamefont
  {Ma}}, \bibinfo {author} {\bibfnamefont {L.}~\bibnamefont {Kou}}, \bibinfo
  {author} {\bibfnamefont {X.}~\bibnamefont {Li}}, \bibinfo {author}
  {\bibfnamefont {Y.}~\bibnamefont {Dai}}, \bibinfo {author} {\bibfnamefont
  {S.~C.}\ \bibnamefont {Smith}}, \ and\ \bibinfo {author} {\bibfnamefont
  {T.}~\bibnamefont {Heine}},\ }\bibfield  {title} {\enquote {\bibinfo {title}
  {Quantum spin hall effect and topological phase transition in two-dimensional
  square transition-metal dichalcogenides},}\ }\href {\doibase
  10.1103/PhysRevB.92.085427} {\bibfield  {journal} {\bibinfo  {journal} {Phys.
  Rev. B}\ }\textbf {\bibinfo {volume} {92}},\ \bibinfo {pages} {085427}
  (\bibinfo {year} {2015})}\BibitemShut {NoStop}%
\bibitem [{\citenamefont {Z.Gong}\ \emph {et~al.}(2013)\citenamefont {Z.Gong},
  \citenamefont {Liu}, \citenamefont {Yu}, \citenamefont {Xiao}, \citenamefont
  {Cui}, \citenamefont {Xu},\ and\ \citenamefont {Yao}}]{Gong}%
  \BibitemOpen
  \bibfield  {author} {\bibinfo {author} {\bibnamefont {Z.Gong}}, \bibinfo
  {author} {\bibfnamefont {G.-B.}\ \bibnamefont {Liu}}, \bibinfo {author}
  {\bibfnamefont {H.}~\bibnamefont {Yu}}, \bibinfo {author} {\bibfnamefont
  {D.}~\bibnamefont {Xiao}}, \bibinfo {author} {\bibfnamefont {X.}~\bibnamefont
  {Cui}}, \bibinfo {author} {\bibfnamefont {X.}~\bibnamefont {Xu}}, \ and\
  \bibinfo {author} {\bibfnamefont {W.}~\bibnamefont {Yao}},\ }\bibfield
  {title} {\enquote {\bibinfo {title} {Magnetoelectric effects and
  valley-controlled spin quantum gates in transition metal dichalcogenide
  bilayers},}\ }\href {\doibase 10.1038/ncomms3053} {\bibfield  {journal}
  {\bibinfo  {journal} {Nat.Commun.}\ }\textbf {\bibinfo {volume} {4}},\
  \bibinfo {pages} {2053} (\bibinfo {year} {2013})}\BibitemShut {NoStop}%
\bibitem [{\citenamefont {Schmidt}\ \emph {et~al.}(2016)\citenamefont
  {Schmidt}, \citenamefont {Yudhistira}, \citenamefont {Chu}, \citenamefont
  {Castro~Neto}, \citenamefont {\"Ozyilmaz}, \citenamefont {Adam},\ and\
  \citenamefont {Eda}}]{Schmidt}%
  \BibitemOpen
  \bibfield  {author} {\bibinfo {author} {\bibfnamefont {H.}~\bibnamefont
  {Schmidt}}, \bibinfo {author} {\bibfnamefont {I.}~\bibnamefont {Yudhistira}},
  \bibinfo {author} {\bibfnamefont {L.}~\bibnamefont {Chu}}, \bibinfo {author}
  {\bibfnamefont {A.~H.}\ \bibnamefont {Castro~Neto}}, \bibinfo {author}
  {\bibfnamefont {B.}~\bibnamefont {\"Ozyilmaz}}, \bibinfo {author}
  {\bibfnamefont {S.}~\bibnamefont {Adam}}, \ and\ \bibinfo {author}
  {\bibfnamefont {G.}~\bibnamefont {Eda}},\ }\bibfield  {title} {\enquote
  {\bibinfo {title} {Quantum transport and observation of dyakonov-perel
  spin-orbit scattering in monolayer ${\mathrm{mos}}_{2}$},}\ }\href {\doibase
  10.1103/PhysRevLett.116.046803} {\bibfield  {journal} {\bibinfo  {journal}
  {Phys. Rev. Lett.}\ }\textbf {\bibinfo {volume} {116}},\ \bibinfo {pages}
  {046803} (\bibinfo {year} {2016})}\BibitemShut {NoStop}%
\bibitem [{\citenamefont {Yang}\ \emph {et~al.}(2015)\citenamefont {Yang},
  \citenamefont {Sinitsyn}, \citenamefont {Chen}, \citenamefont {Yuan},
  \citenamefont {Zhang}, \citenamefont {Lou},\ and\ \citenamefont
  {Crooker}}]{L_Yang}%
  \BibitemOpen
  \bibfield  {author} {\bibinfo {author} {\bibfnamefont {L.}~\bibnamefont
  {Yang}}, \bibinfo {author} {\bibfnamefont {N.~A.}\ \bibnamefont {Sinitsyn}},
  \bibinfo {author} {\bibfnamefont {W.}~\bibnamefont {Chen}}, \bibinfo {author}
  {\bibfnamefont {J.}~\bibnamefont {Yuan}}, \bibinfo {author} {\bibfnamefont
  {J.}~\bibnamefont {Zhang}}, \bibinfo {author} {\bibfnamefont
  {J.}~\bibnamefont {Lou}}, \ and\ \bibinfo {author} {\bibfnamefont {S.~A.}\
  \bibnamefont {Crooker}},\ }\bibfield  {title} {\enquote {\bibinfo {title}
  {Long-lived nanosecond spin relaxation and spin coherence of electrons in
  monolayer mos2 and ws2},}\ }\href {\doibase 10.1038/nphys3419} {\bibfield
  {journal} {\bibinfo  {journal} {Nat. Phyiscs}\ }\textbf {\bibinfo {volume}
  {11}},\ \bibinfo {pages} {830} (\bibinfo {year} {2015})}\BibitemShut
  {NoStop}%
\bibitem [{\citenamefont {Cheng}\ \emph {et~al.}(2013)\citenamefont {Cheng},
  \citenamefont {Zhu}, \citenamefont {Tahir},\ and\ \citenamefont
  {Schwingenschlögl}}]{Cheng}%
  \BibitemOpen
  \bibfield  {author} {\bibinfo {author} {\bibfnamefont {Y.~C.}\ \bibnamefont
  {Cheng}}, \bibinfo {author} {\bibfnamefont {Z.~Y.}\ \bibnamefont {Zhu}},
  \bibinfo {author} {\bibfnamefont {M.}~\bibnamefont {Tahir}}, \ and\ \bibinfo
  {author} {\bibfnamefont {U.}~\bibnamefont {Schwingenschlögl}},\ }\bibfield
  {title} {\enquote {\bibinfo {title} {Spin-orbit–induced spin splittings in
  polar transition metal dichalcogenide monolayers},}\ }\href
  {http://stacks.iop.org/0295-5075/102/i=5/a=57001} {\bibfield  {journal}
  {\bibinfo  {journal} {EPL}\ }\textbf {\bibinfo {volume} {102}},\ \bibinfo
  {pages} {57001} (\bibinfo {year} {2013})}\BibitemShut {NoStop}%
\bibitem [{\citenamefont {Defo}\ \emph {et~al.}(2016)\citenamefont {Defo},
  \citenamefont {Fang}, \citenamefont {Shirodkar}, \citenamefont {Tritsaris},
  \citenamefont {Dimoulas},\ and\ \citenamefont {Kaxiras}}]{Defo}%
  \BibitemOpen
  \bibfield  {author} {\bibinfo {author} {\bibfnamefont {R.~K.}\ \bibnamefont
  {Defo}}, \bibinfo {author} {\bibfnamefont {S.}~\bibnamefont {Fang}}, \bibinfo
  {author} {\bibfnamefont {S.~N.}\ \bibnamefont {Shirodkar}}, \bibinfo {author}
  {\bibfnamefont {G.~A.}\ \bibnamefont {Tritsaris}}, \bibinfo {author}
  {\bibfnamefont {A.}~\bibnamefont {Dimoulas}}, \ and\ \bibinfo {author}
  {\bibfnamefont {E.}~\bibnamefont {Kaxiras}},\ }\bibfield  {title} {\enquote
  {\bibinfo {title} {Strain dependence of band gaps and exciton energies in
  pure and mixed transition-metal dichalcogenides},}\ }\href {\doibase
  10.1103/PhysRevB.94.155310} {\bibfield  {journal} {\bibinfo  {journal} {Phys.
  Rev. B}\ }\textbf {\bibinfo {volume} {94}},\ \bibinfo {pages} {155310}
  (\bibinfo {year} {2016})}\BibitemShut {NoStop}%
\bibitem [{\citenamefont {Xenogiannopoulou}\ \emph {et~al.}(2015)\citenamefont
  {Xenogiannopoulou}, \citenamefont {Tsipas}, \citenamefont {Aretouli},
  \citenamefont {Tsoutsou}, \citenamefont {Giamini}, \citenamefont {Bazioti},
  \citenamefont {Dimitrakopulos}, \citenamefont {Komninou}, \citenamefont
  {Brems}, \citenamefont {Huyghebaert}, \citenamefont {Radu},\ and\
  \citenamefont {Dimoulas}}]{Xenogiannopoulou}%
  \BibitemOpen
  \bibfield  {author} {\bibinfo {author} {\bibfnamefont {E.}~\bibnamefont
  {Xenogiannopoulou}}, \bibinfo {author} {\bibfnamefont {P.}~\bibnamefont
  {Tsipas}}, \bibinfo {author} {\bibfnamefont {K.~E.}\ \bibnamefont
  {Aretouli}}, \bibinfo {author} {\bibfnamefont {D.}~\bibnamefont {Tsoutsou}},
  \bibinfo {author} {\bibfnamefont {S.~A.}\ \bibnamefont {Giamini}}, \bibinfo
  {author} {\bibfnamefont {C.}~\bibnamefont {Bazioti}}, \bibinfo {author}
  {\bibfnamefont {G.~P.}\ \bibnamefont {Dimitrakopulos}}, \bibinfo {author}
  {\bibfnamefont {P.}~\bibnamefont {Komninou}}, \bibinfo {author}
  {\bibfnamefont {S.}~\bibnamefont {Brems}}, \bibinfo {author} {\bibfnamefont
  {C.}~\bibnamefont {Huyghebaert}}, \bibinfo {author} {\bibfnamefont {I.~P.}\
  \bibnamefont {Radu}}, \ and\ \bibinfo {author} {\bibfnamefont
  {A.}~\bibnamefont {Dimoulas}},\ }\bibfield  {title} {\enquote {\bibinfo
  {title} {High-quality{,} large-area mose2 and mose2/bi2se3 heterostructures
  on aln(0001)/si(111) substrates by molecular beam epitaxy},}\ }\href
  {\doibase 10.1039/C4NR06874B} {\bibfield  {journal} {\bibinfo  {journal}
  {Nanoscale}\ }\textbf {\bibinfo {volume} {7}},\ \bibinfo {pages} {7896}
  (\bibinfo {year} {2015})}\BibitemShut {NoStop}%
\bibitem [{\citenamefont {Aretouli}\ \emph {et~al.}(2015)\citenamefont
  {Aretouli}, \citenamefont {Tsipas}, \citenamefont {Tsoutsou}, \citenamefont
  {Marquez-Velasco}, \citenamefont {Xenogiannopoulou}, \citenamefont {Giamini},
  \citenamefont {Vassalou}, \citenamefont {Kelaidis},\ and\ \citenamefont
  {Dimoulas}}]{Aretouli}%
  \BibitemOpen
  \bibfield  {author} {\bibinfo {author} {\bibfnamefont {K.~E.}\ \bibnamefont
  {Aretouli}}, \bibinfo {author} {\bibfnamefont {P.}~\bibnamefont {Tsipas}},
  \bibinfo {author} {\bibfnamefont {D.}~\bibnamefont {Tsoutsou}}, \bibinfo
  {author} {\bibfnamefont {J.}~\bibnamefont {Marquez-Velasco}}, \bibinfo
  {author} {\bibfnamefont {E.}~\bibnamefont {Xenogiannopoulou}}, \bibinfo
  {author} {\bibfnamefont {S.~A.}\ \bibnamefont {Giamini}}, \bibinfo {author}
  {\bibfnamefont {E.}~\bibnamefont {Vassalou}}, \bibinfo {author}
  {\bibfnamefont {N.}~\bibnamefont {Kelaidis}}, \ and\ \bibinfo {author}
  {\bibfnamefont {A.}~\bibnamefont {Dimoulas}},\ }\bibfield  {title} {\enquote
  {\bibinfo {title} {Two-dimensional semiconductor hfse2 and mose2/hfse2 van
  der waals heterostructures by molecular beam epitaxy},}\ }\href {\doibase
  10.1063/1.4917422} {\bibfield  {journal} {\bibinfo  {journal} {Applied
  Physics Letters}\ }\textbf {\bibinfo {volume} {106}},\ \bibinfo {pages}
  {143105} (\bibinfo {year} {2015})}\BibitemShut {NoStop}%
\bibitem [{\citenamefont {Perdew}, \citenamefont {Burke},\ and\ \citenamefont
  {Ernzerhof}(1996)}]{Perdew}%
  \BibitemOpen
  \bibfield  {author} {\bibinfo {author} {\bibfnamefont {J.~P.}\ \bibnamefont
  {Perdew}}, \bibinfo {author} {\bibfnamefont {K.}~\bibnamefont {Burke}}, \
  and\ \bibinfo {author} {\bibfnamefont {M.}~\bibnamefont {Ernzerhof}},\
  }\bibfield  {title} {\enquote {\bibinfo {title} {Generalized gradient
  approximation made simple},}\ }\href {\doibase 10.1103/PhysRevLett.77.3865}
  {\bibfield  {journal} {\bibinfo  {journal} {Phys. Rev. Lett.}\ }\textbf
  {\bibinfo {volume} {77}},\ \bibinfo {pages} {3865--3868} (\bibinfo {year}
  {1996})}\BibitemShut {NoStop}%
\bibitem [{\citenamefont {Ozaki}\ \emph {et~al.}()\citenamefont {Ozaki},
  \citenamefont {Kino}, \citenamefont {Yu}, \citenamefont {Han}, \citenamefont
  {Kobayashi}, \citenamefont {Ohfuti}, \citenamefont {Ishii}, \citenamefont
  {Ohwaki}, \citenamefont {Weng},\ and\ \citenamefont {Terakura}}]{Openmx}%
  \BibitemOpen
  \bibfield  {author} {\bibinfo {author} {\bibfnamefont {T.}~\bibnamefont
  {Ozaki}}, \bibinfo {author} {\bibfnamefont {H.}~\bibnamefont {Kino}},
  \bibinfo {author} {\bibfnamefont {J.}~\bibnamefont {Yu}}, \bibinfo {author}
  {\bibfnamefont {M.~J.}\ \bibnamefont {Han}}, \bibinfo {author} {\bibfnamefont
  {N.}~\bibnamefont {Kobayashi}}, \bibinfo {author} {\bibfnamefont
  {M.}~\bibnamefont {Ohfuti}}, \bibinfo {author} {\bibfnamefont
  {F.}~\bibnamefont {Ishii}}, \bibinfo {author} {\bibfnamefont
  {T.}~\bibnamefont {Ohwaki}}, \bibinfo {author} {\bibfnamefont
  {H.}~\bibnamefont {Weng}}, \ and\ \bibinfo {author} {\bibfnamefont
  {K.}~\bibnamefont {Terakura}},\ }\href@noop {} {}\bibinfo {howpublished}
  {{http://www.openmx-square.org/}}\BibitemShut {NoStop}%
\bibitem [{\citenamefont {Troullier}\ and\ \citenamefont
  {Martins}(1991)}]{Troullier}%
  \BibitemOpen
  \bibfield  {author} {\bibinfo {author} {\bibfnamefont {N.}~\bibnamefont
  {Troullier}}\ and\ \bibinfo {author} {\bibfnamefont {J.~L.}\ \bibnamefont
  {Martins}},\ }\bibfield  {title} {\enquote {\bibinfo {title} {Efficient
  pseudopotentials for plane-wave calculations},}\ }\href {\doibase
  10.1103/PhysRevB.43.1993} {\bibfield  {journal} {\bibinfo  {journal} {Phys.
  Rev. B}\ }\textbf {\bibinfo {volume} {43}},\ \bibinfo {pages} {1993--2006}
  (\bibinfo {year} {1991})}\BibitemShut {NoStop}%
\bibitem [{\citenamefont {Ozaki}(2003)}]{Ozaki}%
  \BibitemOpen
  \bibfield  {author} {\bibinfo {author} {\bibfnamefont {T.}~\bibnamefont
  {Ozaki}},\ }\bibfield  {title} {\enquote {\bibinfo {title} {Variationally
  optimized atomic orbitals for large-scale electronic structures},}\ }\href
  {\doibase 10.1103/PhysRevB.67.155108} {\bibfield  {journal} {\bibinfo
  {journal} {Phys. Rev. B}\ }\textbf {\bibinfo {volume} {67}},\ \bibinfo
  {pages} {155108} (\bibinfo {year} {2003})}\BibitemShut {NoStop}%
\bibitem [{\citenamefont {Ozaki}\ and\ \citenamefont {Kino}(2004)}]{Ozakikino}%
  \BibitemOpen
  \bibfield  {author} {\bibinfo {author} {\bibfnamefont {T.}~\bibnamefont
  {Ozaki}}\ and\ \bibinfo {author} {\bibfnamefont {H.}~\bibnamefont {Kino}},\
  }\bibfield  {title} {\enquote {\bibinfo {title} {Numerical atomic basis
  orbitals from h to kr},}\ }\href {\doibase 10.1103/PhysRevB.69.195113}
  {\bibfield  {journal} {\bibinfo  {journal} {Phys. Rev. B}\ }\textbf {\bibinfo
  {volume} {69}},\ \bibinfo {pages} {195113} (\bibinfo {year}
  {2004})}\BibitemShut {NoStop}%
\bibitem [{\citenamefont {Kotaka}, \citenamefont {Ishii},\ and\ \citenamefont
  {Saito}(2013)}]{Kotaka}%
  \BibitemOpen
  \bibfield  {author} {\bibinfo {author} {\bibfnamefont {H.}~\bibnamefont
  {Kotaka}}, \bibinfo {author} {\bibfnamefont {F.}~\bibnamefont {Ishii}}, \
  and\ \bibinfo {author} {\bibfnamefont {M.}~\bibnamefont {Saito}},\ }\bibfield
   {title} {\enquote {\bibinfo {title} {Rashba effect on the structure of the
  bi one-bilayer film: Fully relativistic first-principles calculation},}\
  }\href {http://stacks.iop.org/1347-4065/52/i=3R/a=035204} {\bibfield
  {journal} {\bibinfo  {journal} {Jpn. J. Appl. Phys.}\ }\textbf {\bibinfo
  {volume} {52}},\ \bibinfo {pages} {035204} (\bibinfo {year}
  {2013})}\BibitemShut {NoStop}%
\bibitem [{\citenamefont {Absor}\ \emph {et~al.}(2014)\citenamefont {Absor},
  \citenamefont {Kotaka}, \citenamefont {Ishii},\ and\ \citenamefont
  {Saito}}]{Absor1}%
  \BibitemOpen
  \bibfield  {author} {\bibinfo {author} {\bibfnamefont {M.~A.~U.}\
  \bibnamefont {Absor}}, \bibinfo {author} {\bibfnamefont {H.}~\bibnamefont
  {Kotaka}}, \bibinfo {author} {\bibfnamefont {F.}~\bibnamefont {Ishii}}, \
  and\ \bibinfo {author} {\bibfnamefont {M.}~\bibnamefont {Saito}},\ }\bibfield
   {title} {\enquote {\bibinfo {title} {Tunable rashba effect on strained zno:
  First-principles density-functional study},}\ }\href
  {http://stacks.iop.org/1882-0786/7/i=5/a=053002} {\bibfield  {journal}
  {\bibinfo  {journal} {Applied Physics Express}\ }\textbf {\bibinfo {volume}
  {7}},\ \bibinfo {pages} {053002} (\bibinfo {year} {2014})}\BibitemShut
  {NoStop}%
\bibitem [{\citenamefont {Absor}\ \emph {et~al.}(2015)\citenamefont {Absor},
  \citenamefont {Ishii}, \citenamefont {Kotaka},\ and\ \citenamefont
  {Saito}}]{Absor2}%
  \BibitemOpen
  \bibfield  {author} {\bibinfo {author} {\bibfnamefont {M.~A.~U.}\
  \bibnamefont {Absor}}, \bibinfo {author} {\bibfnamefont {F.}~\bibnamefont
  {Ishii}}, \bibinfo {author} {\bibfnamefont {H.}~\bibnamefont {Kotaka}}, \
  and\ \bibinfo {author} {\bibfnamefont {M.}~\bibnamefont {Saito}},\ }\bibfield
   {title} {\enquote {\bibinfo {title} {Persistent spin helix on a wurtzite zno
  {$(10\bar{1}0)$} surface: First-principles density-functional study},}\
  }\href {http://stacks.iop.org/1882-0786/8/i=7/a=073006} {\bibfield  {journal}
  {\bibinfo  {journal} {Applied Physics Express}\ }\textbf {\bibinfo {volume}
  {8}},\ \bibinfo {pages} {073006} (\bibinfo {year} {2015})}\BibitemShut
  {NoStop}%
\bibitem [{\citenamefont {Absor}\ \emph
  {et~al.}(2016{\natexlab{b}})\citenamefont {Absor}, \citenamefont {Ishii},
  \citenamefont {Kotaka},\ and\ \citenamefont {Saito}}]{Absor3}%
  \BibitemOpen
  \bibfield  {author} {\bibinfo {author} {\bibfnamefont {M.~A.~U.}\
  \bibnamefont {Absor}}, \bibinfo {author} {\bibfnamefont {F.}~\bibnamefont
  {Ishii}}, \bibinfo {author} {\bibfnamefont {H.}~\bibnamefont {Kotaka}}, \
  and\ \bibinfo {author} {\bibfnamefont {M.}~\bibnamefont {Saito}},\ }\bibfield
   {title} {\enquote {\bibinfo {title} {Spin-split bands of metallic
  hydrogenated zno (101¯0) surface: First-principles study},}\ }\href
  {http://scitation.aip.org/content/aip/journal/adva/6/2/10.1063/1.4942104}
  {\bibfield  {journal} {\bibinfo  {journal} {AIP Advances}\ }\textbf {\bibinfo
  {volume} {6}},\ \bibinfo {eid} {025309} (\bibinfo {year}
  {2016}{\natexlab{b}})}\BibitemShut {NoStop}%
\bibitem [{\citenamefont {Guzman}\ and\ \citenamefont
  {Strachan}(2014)}]{Guzman}%
  \BibitemOpen
  \bibfield  {author} {\bibinfo {author} {\bibfnamefont {D.~M.}\ \bibnamefont
  {Guzman}}\ and\ \bibinfo {author} {\bibfnamefont {A.}~\bibnamefont
  {Strachan}},\ }\bibfield  {title} {\enquote {\bibinfo {title} {Role of strain
  on electronic and mechanical response of semiconducting transition-metal
  dichalcogenide monolayers: An ab-initio study},}\ }\href {\doibase
  10.1063/1.4883995} {\bibfield  {journal} {\bibinfo  {journal} {Journal of
  Applied Physics}\ }\textbf {\bibinfo {volume} {115}},\ \bibinfo {pages}
  {243701} (\bibinfo {year} {2014})}\BibitemShut {NoStop}%
\bibitem [{\citenamefont {Kormányos}\ \emph {et~al.}(2015)\citenamefont
  {Kormányos}, \citenamefont {Burkard}, \citenamefont {Gmitra}, \citenamefont
  {Fabian}, \citenamefont {Zólyomi}, \citenamefont {Drummond},\ and\
  \citenamefont {Fal’ko}}]{AndorB}%
  \BibitemOpen
  \bibfield  {author} {\bibinfo {author} {\bibfnamefont {A.}~\bibnamefont
  {Kormányos}}, \bibinfo {author} {\bibfnamefont {G.}~\bibnamefont {Burkard}},
  \bibinfo {author} {\bibfnamefont {M.}~\bibnamefont {Gmitra}}, \bibinfo
  {author} {\bibfnamefont {J.}~\bibnamefont {Fabian}}, \bibinfo {author}
  {\bibfnamefont {V.}~\bibnamefont {Zólyomi}}, \bibinfo {author}
  {\bibfnamefont {N.~D.}\ \bibnamefont {Drummond}}, \ and\ \bibinfo {author}
  {\bibfnamefont {V.}~\bibnamefont {Fal’ko}},\ }\bibfield  {title} {\enquote
  {\bibinfo {title} {k · p theory for two-dimensional transition metal
  dichalcogenide semiconductors},}\ }\href
  {http://stacks.iop.org/2053-1583/2/i=2/a=022001} {\bibfield  {journal}
  {\bibinfo  {journal} {2D Materials}\ }\textbf {\bibinfo {volume} {2}},\
  \bibinfo {pages} {022001} (\bibinfo {year} {2015})}\BibitemShut {NoStop}%
\bibitem [{\citenamefont {Brumme}, \citenamefont {Calandra},\ and\
  \citenamefont {Mauri}(2015)}]{Brumme}%
  \BibitemOpen
  \bibfield  {author} {\bibinfo {author} {\bibfnamefont {T.}~\bibnamefont
  {Brumme}}, \bibinfo {author} {\bibfnamefont {M.}~\bibnamefont {Calandra}}, \
  and\ \bibinfo {author} {\bibfnamefont {F.}~\bibnamefont {Mauri}},\ }\bibfield
   {title} {\enquote {\bibinfo {title} {First-principles theory of field-effect
  doping in transition-metal dichalcogenides: Structural properties, electronic
  structure, hall coefficient, and electrical conductivity},}\ }\href {\doibase
  10.1103/PhysRevB.91.155436} {\bibfield  {journal} {\bibinfo  {journal} {Phys.
  Rev. B}\ }\textbf {\bibinfo {volume} {91}},\ \bibinfo {pages} {155436}
  (\bibinfo {year} {2015})}\BibitemShut {NoStop}%
\bibitem [{\citenamefont {Bertolazzi}, \citenamefont {Brivio},\ and\
  \citenamefont {Kis}(2011)}]{Bertolazzi}%
  \BibitemOpen
  \bibfield  {author} {\bibinfo {author} {\bibfnamefont {S.}~\bibnamefont
  {Bertolazzi}}, \bibinfo {author} {\bibfnamefont {J.}~\bibnamefont {Brivio}},
  \ and\ \bibinfo {author} {\bibfnamefont {A.}~\bibnamefont {Kis}},\ }\bibfield
   {title} {\enquote {\bibinfo {title} {Stretching and breaking of ultrathin
  mos2},}\ }\href {\doibase 10.1021/nn203879f} {\bibfield  {journal} {\bibinfo
  {journal} {ACS Nano}\ }\textbf {\bibinfo {volume} {5}},\ \bibinfo {pages}
  {9703--9709} (\bibinfo {year} {2011})},\ \bibinfo {note} {pMID:
  22087740}\BibitemShut {NoStop}%
\bibitem [{\citenamefont {Nilsson}, \citenamefont {Janzén},\ and\
  \citenamefont {Kakanakova-Georgieva}(2016)}]{Nilsson}%
  \BibitemOpen
  \bibfield  {author} {\bibinfo {author} {\bibfnamefont {D.}~\bibnamefont
  {Nilsson}}, \bibinfo {author} {\bibfnamefont {E.}~\bibnamefont {Janzén}}, \
  and\ \bibinfo {author} {\bibfnamefont {A.}~\bibnamefont
  {Kakanakova-Georgieva}},\ }\bibfield  {title} {\enquote {\bibinfo {title}
  {Lattice parameters of aln bulk, homoepitaxial and heteroepitaxial
  material},}\ }\href {http://stacks.iop.org/0022-3727/49/i=17/a=175108}
  {\bibfield  {journal} {\bibinfo  {journal} {Journal of Physics D: Applied
  Physics}\ }\textbf {\bibinfo {volume} {49}},\ \bibinfo {pages} {175108}
  (\bibinfo {year} {2016})}\BibitemShut {NoStop}%
\bibitem [{\citenamefont {Fu}(2009)}]{Fu}%
  \BibitemOpen
  \bibfield  {author} {\bibinfo {author} {\bibfnamefont {L.}~\bibnamefont
  {Fu}},\ }\bibfield  {title} {\enquote {\bibinfo {title} {Hexagonal warping
  effects in the surface states of the topological insulator
  ${\mathrm{bi}}_{2}{\mathrm{te}}_{3}$},}\ }\href {\doibase
  10.1103/PhysRevLett.103.266801} {\bibfield  {journal} {\bibinfo  {journal}
  {Phys. Rev. Lett.}\ }\textbf {\bibinfo {volume} {103}},\ \bibinfo {pages}
  {266801} (\bibinfo {year} {2009})}\BibitemShut {NoStop}%
\bibitem [{\citenamefont {Vajna}\ \emph {et~al.}(2012)\citenamefont {Vajna},
  \citenamefont {Simon}, \citenamefont {Szilva}, \citenamefont {Palotas},
  \citenamefont {Ujfalussy},\ and\ \citenamefont {Szunyogh}}]{Vajna}%
  \BibitemOpen
  \bibfield  {author} {\bibinfo {author} {\bibfnamefont {S.}~\bibnamefont
  {Vajna}}, \bibinfo {author} {\bibfnamefont {E.}~\bibnamefont {Simon}},
  \bibinfo {author} {\bibfnamefont {A.}~\bibnamefont {Szilva}}, \bibinfo
  {author} {\bibfnamefont {K.}~\bibnamefont {Palotas}}, \bibinfo {author}
  {\bibfnamefont {B.}~\bibnamefont {Ujfalussy}}, \ and\ \bibinfo {author}
  {\bibfnamefont {L.}~\bibnamefont {Szunyogh}},\ }\bibfield  {title} {\enquote
  {\bibinfo {title} {Higher-order contributions to the rashba-bychkov effect
  with application to the bi/ag(111) surface alloy},}\ }\href {\doibase
  10.1103/PhysRevB.85.075404} {\bibfield  {journal} {\bibinfo  {journal} {Phys.
  Rev. B}\ }\textbf {\bibinfo {volume} {85}},\ \bibinfo {pages} {075404}
  (\bibinfo {year} {2012})}\BibitemShut {NoStop}%
\bibitem [{\citenamefont {Lee}, \citenamefont {Yun},\ and\ \citenamefont
  {Lee}(2015)}]{Lee}%
  \BibitemOpen
  \bibfield  {author} {\bibinfo {author} {\bibfnamefont {K.}~\bibnamefont
  {Lee}}, \bibinfo {author} {\bibfnamefont {W.~S.}\ \bibnamefont {Yun}}, \ and\
  \bibinfo {author} {\bibfnamefont {J.~D.}\ \bibnamefont {Lee}},\ }\bibfield
  {title} {\enquote {\bibinfo {title} {Giant rashba-type splitting in
  molybdenum-driven bands of ${\mathrm{mos}}_{2}/\mathrm{Bi}(111)$
  heterostructure},}\ }\href {\doibase 10.1103/PhysRevB.91.125420} {\bibfield
  {journal} {\bibinfo  {journal} {Phys. Rev. B}\ }\textbf {\bibinfo {volume}
  {91}},\ \bibinfo {pages} {125420} (\bibinfo {year} {2015})}\BibitemShut
  {NoStop}%
\bibitem [{\citenamefont {Liu}\ \emph {et~al.}(2015)\citenamefont {Liu},
  \citenamefont {Chen}, \citenamefont {Yu}, \citenamefont {Yang}, \citenamefont
  {Jiao}, \citenamefont {Liu}, \citenamefont {Ho}, \citenamefont {Gao},
  \citenamefont {Jia}, \citenamefont {Yao},\ and\ \citenamefont {Xie}}]{Liu}%
  \BibitemOpen
  \bibfield  {author} {\bibinfo {author} {\bibfnamefont {H.}~\bibnamefont
  {Liu}}, \bibinfo {author} {\bibfnamefont {J.}~\bibnamefont {Chen}}, \bibinfo
  {author} {\bibfnamefont {H.}~\bibnamefont {Yu}}, \bibinfo {author}
  {\bibfnamefont {F.}~\bibnamefont {Yang}}, \bibinfo {author} {\bibfnamefont
  {L.}~\bibnamefont {Jiao}}, \bibinfo {author} {\bibfnamefont {G.-B.}\
  \bibnamefont {Liu}}, \bibinfo {author} {\bibfnamefont {W.}~\bibnamefont
  {Ho}}, \bibinfo {author} {\bibfnamefont {C.}~\bibnamefont {Gao}}, \bibinfo
  {author} {\bibfnamefont {J.}~\bibnamefont {Jia}}, \bibinfo {author}
  {\bibfnamefont {W.}~\bibnamefont {Yao}}, \ and\ \bibinfo {author}
  {\bibfnamefont {M.}~\bibnamefont {Xie}},\ }\bibfield  {title} {\enquote
  {\bibinfo {title} {Observation of intervalley quantum interference in
  epitaxial monolayer tungsten diselenide},}\ }\href {\doibase
  10.1038/ncomms9180} {\bibfield  {journal} {\bibinfo  {journal} {Nat. Commun}\
  }\textbf {\bibinfo {volume} {6}},\ \bibinfo {pages} {8180} (\bibinfo {year}
  {2015})}\BibitemShut {NoStop}%
\end{thebibliography}%


\end{document}